\input{psfig.sty}
\input epsf 
\documentstyle[12pt]{article}
\newcommand{\journal}[4]{{\em #1~}#2\,(19#3)\,#4;}

\newcommand{\pr}{\journal {Phys. Rev.}}

\newcommand{\rmp}{\journal {Rev. Mod. Phys.}} 
\newcommand{\cmp}{\journal {Comm. Math. Phys.}}

\newcommand{\np}{\journal {Nucl. Phys.}} 
\newcommand{\pl}{\journal {Phys. Lett.}} 
 
\newcommand{\prep}{\journal {Phys. Reports}} 
 
\newcommand{\nc}{\journal {Nuovo Cim.}}

\newcommand{\annp}{\journal {Ann. Phys. (N.Y.)}}

\makeatletter 
\@addtoreset{equation}{section} 
\makeatother 
 
\newcommand{\ts}{\tilde{s}} 
\renewcommand{\a}{\alpha} 
\renewcommand{\b}{\beta} 
\renewcommand{\d}{\delta}  
\newcommand{\e}{\varepsilon} 
\newcommand{\f}{\phi} 
\newcommand{\g}{\gamma} \newcommand{\G}{\Gamma} 
\newcommand{\k}{\kappa} 
\renewcommand{\L}{\Lambda}
\newcommand{\lm}{\lambda} 
 
\newcommand{\m}{\mu} 
\newcommand{\n}{\nu} 
\newcommand{\mn}{{\m\n}} 
\renewcommand{\o}{\omega}  
\newcommand{\p}{\psi} 
\newcommand{\pb}{\bar\psi} 
\newcommand{\r}{\rho}

\newcommand{\s}{\sigma} \renewcommand{\S}{\Sigma} 
\newcommand{\rs}{{\r\s}} 
\renewcommand{\t}{\theta}

\newcommand{\CC}{{\cal C}} 
\newcommand{\DD}{{\cal D}} 
\newcommand{\EE}{{\cal E}} 
\newcommand{\FF}{{\cal F}}

\newcommand{\MM}{{\cal M}} 
\newcommand{\NN}{{\cal N}} 
\newcommand{\OO}{{\cal O}} 
\newcommand{\PP}{{\cal P}} 
 
\newcommand{\RR}{{\cal R}} 
\newcommand{\TT}{{\cal T}} 
 
\newcommand{\VV}{{\cal V}}

\newcommand{\ZZ}{{\cal Z}} 
 
\newcommand{\complex}{{\kern .1em {\raise .47ex 
\hbox {$\scriptscriptstyle |$}} 
\kern -.4em {\rm C}}} 
\newcommand{\real}{{{\rm I} \kern -.19em {\rm R}}} 
\newcommand{\rational}{{\kern .1em {\raise .47ex 
\hbox{$\scripscriptstyle |$}} 
\kern -.35em {\rm Q}}} 
\renewcommand{\natural}{{\vrule height 1.6ex width 
.05em depth 0ex \kern -.35em {\rm N}}}
\newcommand{\integers}{{\bf Z}} 
\newcommand{\ipr}{\!\cdot\!}

\newcommand{\intd}[1]{\int\frac{d^D#1}{(2\pi)^D}\,}
\newcommand{\intf}[1]{\int\frac{d^4#1}{(2\pi)^4}\,}
\newcommand{\tr}{{\rm {Tr} \,}} 
 
\renewcommand{\exp}{{\rm \ {exp}\,}} 
\newcommand{\cb}{{\bar c}}

\newcommand{\half}{\frac 1 2} 
 
\newcommand{\pa}{\partial} 
\newcommand{\pad}[2]{{\frac{\partial #1}{\partial #2}}} 
\newcommand{\fud}[2]  {{\displaystyle{\frac{\delta #1}{\delta #2}}}}

\newcommand{\sla}{\raise.15ex\hbox{$/$}\kern -.8em} 
 
\newcommand{\twiddle}{\lower.9ex\rlap{$\kern -.1em\scriptstyle\sim$}}

\newcommand{\vev}[1]{\left\langle {#1}\right\rangle}

\newcommand{\equ}[1]{~(\ref{#1})}

\newcommand{\eq}{\begin{equation}} 
\newcommand{\eqn}[1]{\label{#1}\end{equation}} 
\newcommand{\eea}{\end{eqnarray}} 
\newcommand{\eqa}{\begin{eqnarray}} 
\newcommand{\eqan}[1]{\label{#1}\end{eqnarray}} 
\newcommand{\ba}[1]{\begin{equation}\begin{array}{#1}} 
\newcommand{\ea}[1]{\end{array}\label{#1}\end{equation}} 
\newcommand{\eqac}{\begin{equation}\begin{array}{rcl}} 
\newcommand{\eqacn}[1]{\end{array}\label{#1}\end{equation}}

\renewcommand{\pad}[2]{{\displaystyle{\frac{\partial #1}{\partial #2}}}}

\newcommand{\one}{{\bf 1}} 
\renewcommand{\sb}{{\bar \s}} 
\newcommand{\gb}{{\bar \g}} 
 
\begin{document} 
\def\ftoday{{\sl  \number\day \space\ifcase\month  
\or Janvier\or F\'evrier\or Mars\or avril\or Mai 
\or Juin\or Juillet\or Ao\^ut\or Septembre\or Octobre 
\or Novembre \or D\'ecembre\fi 
\space  \number\year}}     
\titlepage 
%
{
\begin{center} 
{ \huge    Confinement in Covariant Gauges} 
\vspace{2ex} 
 
{\Large Martin Schaden\footnote{e-mail~: ms68@scires.nyu.edu\\ 
\indent{~~Research} supported in part by the National Science Foundation 
under grant no.~PHY93-18781} and Alexander Rozenberg\footnote{e-mail~:sasha.rozenberg@nyu.edu}}\\  
{\it\large Physics Department, New York University,\\ 4 Washington Place, 
New York, N.Y. 10003} 
\end{center} 
\vspace{4ex} 
 
\begin{center} 
\bf ABSTRACT 
\end{center} 
 
We examine the weak coupling limit of  Euclidean $SU(n)$ gauge theory
in covariant gauges. Following an   
earlier suggestion, an equivariant BRST-construction is used  to 
define the continuum theory on a finite torus.  The equivariant gauge
fixing introduces constant  ghost fields as moduli of the model. We
study the parameter- and  
moduli- space perturbatively. For $n_f \leq n$ quark
flavors, the moduli flow to a non-trivial fixed point in certain
critical covariant gauges and the one-loop effective potential indicates
that the global $SU(n)$ color symmetry of the gauge fixed model is
spontaneously 
broken to $U(1)^{n-1}$. Ward identities and renormalization
group arguments imply that 
the  longitudinal  gauge boson propagator at long range is dominated
by $n(n-1)$ Goldstone bosons in these critical
covariant gauges. In the large $n$ limit, we derive  a nonlinear
integral equation for the expectation value of large Wilson loops assuming
that the exchange of Goldstone bosons dominates the interaction at
long range in critical covariant gauges. We find numerically that the
expectation value of large circular Wilson loops decreases
exponentially with the enclosed area in the absence of dynamical
fermions. The
gauge invariance of this  mechanism for confinement in critical 
covariant gauges is discussed.

PACS: 12.38.Aw 12.38.Lg 11.15.Bt 11.30.Qc\hfill\break 
NYU-TH--97/06/30\hfil June 1997 
\vfill 
 
\newpage 
\def\be{\begin{eqnarray}} 
\def\ee{\end{eqnarray}} 
\def\nn{\nonumber} 
\section{Introduction}
Some exact results for the  measure on the moduli space of
supersymmetric theories have recently been obtained. Holomorphicity
and the exactness of the one-loop $\beta$-function 
greatly restrict the low energy effective action of these models,
and one can reach exact conclusions about their phase structure. When
the number of matter fields is not too large, these non-abelian
theories confine\cite{se94}.

The absence of fundamental scalars is one of the obstacles which an
investigation of nonperturbative effects in unbroken ordinary gauge
theories has to confront. The moduli parameters in 
this case represent expectation values of
composite operators and there is apparently  no analog to
the powerful constraint of holomorphicity for the effective
potential. It was only recently observed that a  
translationally invariant quantization of ordinary gauge theories on
compact manifolds does introduce moduli in a natural way. To eliminate
normalizable zero modes of the ghosts that would lead to a vanishing
partition function, the global $SU(n)$ symmetry of the 
gauge theory is treated equivariantly in a quantization which manifestly
preserves the isometries of a compact space-time manifold.  This 
equivariant BRST quantization\cite{bs97} 
introduces a number of global moduli, one of which is a scalar  in the
adjoint representation with vanishing ghost number. A consistent
covariant gauge fixing procedure thus introduces global moduli
parameters even if the gauge symmetry is not broken and fundamental
scalars are absent at the classical level. 

We here investigate in detail the moduli-space of the covariantly
quantized $SU(n)$ gauge theory on a torus and study the
phase space of its unphysical degrees of freedom as a function of the
gauge fixing parameters. We
consider the most general renormalizable covariant gauges. They depend on three
moduli and two gauge parameters: in addition to the usual covariant
gauge parameter $\a$ related to the
longitudinal gluon propagator, we also use the freedom to introduce a
quartic ghost coupling in the gauge fixing
functional\cite{ba82}. Although we cannot compute the exact effective
potential, the
perturbative analysis reveals fixed points of the parameter and moduli
space. These indicate that there 
are at least two asymptotically free phases of the  unphysical degrees of
freedom for $n_f\leq n$  quark  flavors. Although no local physical
order parameter distinguishes between the phases, we argue that a
perturbative analysis of the model is consistent only {\it near} the phase
transition. Since the moduli are introduced to avoid infrared problems
in the gauge fixed action, the measure on this
moduli space is gauge dependent. We however argue that it reflects
non-perturbative properties of the gauge fixed model at fixed points
in the parameter space. 

The one-loop effective potential on the moduli space indicates that
the phase transition 
is characterized by a condensation of ghost-antighost pairs in 
the adjoint representation. In the infinite volume limit this 
leads to a spontaneous breaking of the global $SU(n)$-symmetry to
$U(1)^{n-1}$ of the equivariantly gauge fixed model. Although this
pattern of spontaneous symmetry breaking is 
reminiscent of monopole condensation, the composite local scalar field in
this case is gauge dependent and {\it not} a physical order parameter.
One has to consider non-local Wilson\cite{wi74}-  
or equivalently, magnetic\cite{tH79}- loop  operators as gauge
invariant order parameters for confinement.

The scale of
symmetry breaking is found to be RG-invariant and 
proportional to $\Lambda_{ASP}$ on a line in the  parameter
space. For a fixed number $n_f<n$ fermions in the fundamental
representation, there are two fixed points on this line which merge
for $n_f=n$.  The
corresponding covariant gauges we call critical covariant gauges
(CCG), since perturbation 
theory appears to be selfconsistent in these gauges {\it also for 
  unphysical degrees of freedom}.   
 
Ward identities are used to show that the correlator of the composite
local  operator $\OO^a(x)\propto f^{abc}\cb^b(x) c^c(x)$ with the conserved
color current in CCG has a Goldstone pole in the thermodynamic limit.
We infer from this, that $\vev{\pa\ipr A(x)\ \pa\ipr
A(0)}$ also has Goldstone behavior at large distances. This behavior
is shown to be compatible with the Ward identities in CCG. In these gauges
with quartic ghost interactions, Ward identities do not completely
determine the longitudinal gluon propagator, but rather relate it to
correlation functions in the ghost-antighost channel.

The contribution of the long-range longitudinal interaction to
 the expectation value of large Wilson loops is then obtained in leading
 order of $n$. Numerical solution of the non-linear
 integro-differential equation  
 we derive in this approximation shows that the expectation
 value of large circular Wilson loops decreases
 exponentially with the enclosed area over 12  orders of magnitude.

The structure of this article is as follows. In section~2 we obtain
the continuum action of the equivariantly quantized theory in general
covariant gauges on a torus. This step is crucial, since it gauge
fixes also the long wavelength excitations of the Yang-Mills theory
without destroying the homogeneity and isotropy of the model. The
equivariant construction introduces global fields, which we
subsequently  treat as
moduli. In section~3 we study the renormalization flow behavior of the
gauge parameters in such generalized
covariant gauges. We locate the 
ultra-violet fixed point of the gauge parameter $\a$, which is related to the
longitudinal gluon propagator at large momentum transfers. In
section~4 we  determine the non-trivial fixed points of the moduli-space 
introduced by the equivariant gauge
fixing. We find that the minimum of the  effective potential is
at a non-trivial value for one of these
moduli and that the scale of this minimum is proportional
to $\Lambda_{ASP}$ in CCG. The critical exponent of the non-trivial moduli 
vanishes in these critical gauges.  The  one-loop
effective potential also indicates that the global $SU(n)$-symmetry is broken
to $U(1)^{n-1}$ in critical gauges. In section~5 we exploit the
consequences of this symmetry breakdown using  Ward identities of the
model. These provide the connection between the perturbative
indication that the $SU(n)$-symmetry of the model is spontaneously
broken and long-range effects. We establish the
existence of $n(n-1)$  unphysical Goldstone bosons associated with the
global symmetry breaking by studying the $\OO^a$-correlator with the conserved
color current. Due to mixing between $\OO^a$ and $\pa\ipr A$ by
renormalization, the Goldstone bosons also dominate the 
long-range behavior of the longitudinal gluon-propagator in CCG. We
show that this behavior is
consistent with the Ward identities. In section~6 we use these results
to approximately evaluate the
expectation value of large circular Wilson loops in leading order of
$n$. We derive a non-linear integral equation which resums the
longitudinal  interaction in leading order of $n$ and we solve it
numerically.  Section~7 is a summary and discussion of the results.
 
\section{$SU(n)$ gauge theory in covariant gauges} 
To investigate phases of a quantum field theory, one has to 
control  it in the  thermodynamic limit of infinite space-time 
volume.  We will study 
order parameters of an Euclidean  $SU(n)$ gauge theory on a torus as the
space-time volume of the torus becomes large. 

A lattice regularization might appear to be the only
reasonable definition of a gauge theory, since it does not depend on
perturbation theory. Although this is a perfectly regularized 
statistical  theory on a point manifold, the physical volume 
of any finite lattice shrinks exponentially as the coupling 
constant is tuned to the critical value. There is little 
hope of studying long-range effects numerically in the critical 
regime, and current simulations are restricted to a few  fm$^4$ of
physical volume. These numerical investigations suggest 
confinement, but deconfinement  at a couple of  GeV excitation energy
cannot be  excluded by these  studies. Experimental evidence for  the
absence of  colored  (and/or fractionally charged) asymptotic states is 
considerably  more restrictive and likely to remain so for some time. 

A theoretical investigation of the thermodynamic limit of  
lattice gauge theory near its critical point is rather
complicated. It has  
been pointed out\cite{zw94}, that the Gribov problem\cite{gr78} cannot
be ignored in a study of the lattice in this critical limit. A proposed 
effective description of  lattice gauge theory
at weak coupling introduces an additional parameter\cite{zw94}, which
effectively constrains the   
configuration space to the fundamental modular region in the
thermodynamic limit. It was shown\cite{ms94} that a non-vanishing  
value for this parameter spontaneously breaks the
BRST-symmetry of a lattice in Landau gauge\footnote{The BRST-symmetry
of a covariantly fixed {\it 
finite} lattice has to be broken, since the Euler character of the
lattice gauge group vanishes and the gauge fixed partition function of
any BRST-symmetric finite lattice model would therefore vanish as
well\cite{brs96}.}.
 
We will not attempt to derive the thermodynamic limit of 
a lattice gauge theory here, but propose to investigate  instead the
consistency  of perturbation theory in a covariantly gauge fixed
continuum theory with presumably the same critical behavior as a
lattice gauge theory. 
Let us first perturbatively  define the  continuum gauge theory 
on a compact space-time manifold such as a hypertorus. In order to
effectively exploit the space-time symmetries of the thermodynamic
limit, we consider only gauges which manifestly preserve {\it all} the
isometries of the torus, i.e. general covariant gauges. The 
torus with periodic boundary conditions for the gauge and ghost fields
is chosen for the following reasons:   
\begin{itemize} 
\item analytic continuation  in the dimension $D$ of the hypertorus 
$T_D$ of perturbative expressions is  
 straightforward. The perturbative dimensional regularization does 
not break the gauge- (or better BRST-) invariance explicitly. It
allows one to efficiently extract the ultraviolet  finite renormalized
quantities 
in the limit $D\rightarrow 4$ and  their critical exponents. 
\item The space of gauge  {\it orbits} of an $SU(n)$ theory on a
hypertorus with periodic boundary conditions is
connected.  We thus only 
have to consider perturbation theory around a {\it  single} 
classical gauge orbit.  The absence of nontrivial topological sectors 
on the finite torus with periodic boundary conditions might   
oversimplify our study of the thermodynamic limit of this model. 
The non-perturbative effects we wish to
investigate, however seem to be more easily described in a toroidal 
geometry of Euclidean space-time.  
\item An $SU(n)$ gauge theory defined
on a torus resolves the topological ambiguities of  the covariant
gauge fixing procedure we 
employ\cite{bs97}. These topological anomalies  are present only on base
manifolds with non-trivial 3-cycles. 
\item there is no topological obstruction to introducing
  fermions\cite{ns83} in the fundamental representation on a torus
  with periodic boundary conditions. This would not
  be possible with twisted boundary conditions for the gauge
  field\cite{tH79}. 
\item The torus is a flat Euclidean space-time manifold.  Translations
and the associated Fourier - analysis are straightforward, and an explicit
mode expansion is manageable.
\end{itemize} 
These  topological properties  of a torus with periodic boundary
conditions facilitate the perturbative
analysis considerably. We will repeatedly  take advantage of 
the translational and rotational invariance of gauge-dependent correlation 
functions in the thermodynamic limit of infinite  space-time
volume in covariant gauges.   

\subsection{Equivariant gauge fixing} 
Thermodynamic arguments for {\it gauge dependent} correlation
functions are valid in {\it covariant} gauges, which 
by definition preserve the isometries of the base manifold. The
Goldstone theorem  in this case also gives information about the
long-range behavior of {\it  gauge-dependent} Green's functions. 
  
It is well known that harmonic ghost modes pose an 
obstacle to the definition of a covariantly  gauge-fixed  partition
function on compact manifolds\cite{mi81}. These zero-modes of the
Faddeev-Popov ghosts are  
a consequence of the invariance of covariant gauges 
with respect to {\it global} $SU(n)$ rotations. The zero modes are
conventionally removed by 
fixing the global $SU(n)$ invariance at  a particular space-time 
point\cite{mi81}. Although this procedure preserves the BRST-algebra, 
such pointed gauges single out a space-time point, 
and thus break the translational invariance of gauge-dependent 
correlators explicitly.  If an unbroken 
$SU(n)$ gauge theory indeed confines color, the usual argument that this 
point-defect 
can be disregarded in the thermodynamic limit of gauge dependent
correlation functions  does not necessarily hold,  
since color correlations must be strong and of long range. It is 
therefore not obvious that translational invariance of gauge 
dependent Green's functions, such as gluon- or ghost-propagators, is restored
in the thermodynamic limit of   
pointed gauges.  If translational and Euclidean rotational
invariance  of gauge dependent
correlators on the other hand cannot be explicitly maintained, the
motivation for  choosing a covariant gauge becomes questionable.
 
Recently a solution to the problem with harmonic ghost
zero modes was proposed, which manifestly preserves the invariance of
the gauge fixed theory with respect to {\it all}  isometries of
a compact
manifold\cite{bs97}. The global $SU(n)$-invariance is treated
equivariantly in this construction and the usual field content of 
an $SU(n)$ gauge theory augmented by constant global ghosts whose 
canonical dimensions and ghost numbers are given in Table~1. 
\begin{center} 
\begin{tabular}{|l|r|r||r|r|r|r|r|r|r|r|r|}\hline 
field&$A_\m(x)$ & $\p_i(x)\&\pb_i(x)$ & $c(x)$ & $\cb(x)$ & $b(x)$ & $\f$ & 
$\o$ & $\s$ & $\sb$ & $\gb$ & $\g$\\ \hline 
dim&$1$&$3/2$&$0$&$2$&$2$&$0$&$0$&$4$&$4$&$2$&$2$\\ \hline 
$\f\Pi$&$0$&$0$&$1$&$-1$&$0$&$2$&$1$&$-2$&$-1$&$0$&$1$\\ \hline 
\end{tabular} 
 
\nobreak\vspace{.2cm}{\footnotesize 
{\bf Table 1.} Dimensions and ghost numbers of the fields.}  
\end{center} 
The action of the  
nilpotent BRST-operator $s$ on the fields of Table~1 is\footnote{We will 
often suppress color indices. Except for $\p_i,\pb_i$, vectors of 
the two fundamental representations of  $SU(n)$,  all fields are 
traceless, anti-hermitian $n\times n$ matrices in this notation,  and 
transform under the adjoint representation.}  
\be\label{sdef} 
s A_\mu(x) &=& D^{A }_\mu c(x)-[\o, A _\mu(x)]\nn\\ 
s c(x) &=& -[\o, c(x)] -\frac{g}{2}[ c (x), c (x)] -\frac{1}{g}\f\nn\\ 
s \o &=& -\half [ \o,\o] +\f\nn\\ 
s \f &=& -[\o, \f]\nn\\ 
s \cb(x)&=&-[\o,\cb(x)] + b(x)\nn\\ 
s b(x) &=& -[\o, b(x)] + [\f, \cb(x)]\nn\\ 
s \s &=& -[\o,\s]  +\sb \nn\\ 
s \sb &=& -[\o, \sb] + [\f, \s]\nn\\ 
s \gb &=& -[\o,\gb] +\g\nn\\ 
s \g &=& -[\o, \g] + [\f, \gb]\nn\\ 
s \p_i(x) &=&-\o\p_i(x) -g c(x)\p_i(x) \nn\\  
s \pb_i(x) &=& -\pb_i(x)\o -g\pb_i(x) c(x) 
\ee 
where 
\eq 
D^A_\m c(x) = \pa_\m c(x) + g[A_\m(x), c(x)] 
\eqn{covderivative} 
is the usual covariant derivative for the adjoint representation.  
 
It is straightforward to show that the BRST-operator defined above is  
nilpotent  on any element of the graded algebra constructed from the 
fields of Table~1: 
\eq 
s^2=0 
\eqn{nilpotency} 
The constant ghost $\o$ generates global  gauge 
transformations of all the 
fields in the BRST algebra, except itself. One therefore can
restrict observables to the equivariant cohomology     
$\S$, 
\eq 
\S =\{\OO   :\pad{\OO}{\o^a}=0; s\OO=0, \OO\neq s\FF\} 
\eqn{observables} 
where $\FF$ is itself $\o$-independent.  Since one is 
interested only  in expectation values of gauge invariant functionals 
of $A,\pb $ and $\p$, the notion of {\it 
physical} observables can be further sharpened to functionals in 
the equivariant  cohomology with vanishing ghost number\footnote{To avoid 
global anomalies due to {\it large} gauge transformations, one could
consider only {\it  
local} observables. However, the Chern-Simons term appears to be the only {\it
gobal} physical observable with such a gauge-
ambiguity\cite{bs97}. The restriction to {\it local} observables is therefore
not necessary on a torus.}.  
The most general power counting renormalizable classical  action $S$ with the 
symmetry \equ{sdef} depends on two gauge parameters $\a$ and $\b=\a\d$, 
\eq  
S= S_C [g, m_i; A,\p_i,\pb_i]+ s W_{GF}[g,\a,\b,\gb,\s;A,\cb,c] 
\eqn{action} 
where $S_C$ is the Yang-Mills action of an $SU(n)$ gauge theory with 
$n_f$ quark flavors in the fundamental representation   
\eq 
S_C = \int_\TT \half \tr F_\mn(A) F_\mn(A)+ \sum_{i=1}^{n_f} \pb_i 
(\sla D +m_i)\p_i  
\eqn{classaction} 
The Euclidean Dirac operator here is\footnote{The Euclidean Dirac 
matrices $\g_\m$ satisfy  $\g_\m\g_\n +\g_\n\g_\m =2\d_\mn \one$.} 
\eq 
\sla D\p_i=\g_\m (\pa_\m +g A_\m)\p_i 
\eqn{dirac} 
and  the field strength is related to the connection by
\eq 
F_\mn(A) = \pa_\m A_\n-\pa_\n A_\m +g[A_\m, A_\n] 
\eqn{fieldstrength} 
Note that a CP-breaking term proportional to the Pontryagin number 
$\int_\TT \tr F\wedge F$ vanishes on a torus with periodic boundary
conditions\footnote{``Toron'' sectors with
non-vanishing Pontryagin number exist on a torus with
twisted boundary conditions, which however can only be imposed in the
absence of quarks in the fundamental representation\cite{tH79}.}. We therefore
do not include such a term in\equ{classaction}. The discrete
CP-symmetry in this sense is not broken on any finite torus.   
 
The BRST-exact gauge fixing term  of the action\equ{action} was 
obtained in\cite{bs97}. 
\eq 
W_{GF} =  2\int_\TT dx\ \tr [\pa^\mu \cb(x) A_\mu(x) -
\half\a\cb(x)b(x)-\a\d g \cb(x)\cb(x)c(x)  
+\frac{1}{g}\gb  \cb(x) + \frac{1}{g}\s  c(x)] 
\eqn{wgf} 
For $\a>0$ the Nakanishi-Lautrup field $b(x)$ in\equ{action} can 
be eliminated using the equation of motion 
\eq 
b(x)=(\gb/g -\pa\ipr A(x))/\a -\d g[\cb(x), c(x)], 
\eqn{EMb} 
since observables do not depend on $b(x)$.

A covariantly quantized $SU(n)$ gauge theory on a torus is thus
described by the tree-level action  
\be\label{effaction} 
S_0&=&S_C + 2\int_\TT dx\ \tr\left[\frac{1}{2\a}(\pa\ipr A(x))^2 + 
\d\cb(x)\, D^A\ipr\pa c(x) + (1-\d)\cb(x)\pa\ipr D^A c(x) + 
\right.\nn\\ 
&&\qquad +\a\d(1-\d)g^2\cb(x)\cb(x) c(x) c(x)-\d [\cb(x),c(x)]\gb 
+\frac{1}{2\a g^2}\gb^2 -\frac{1}{\a g}\gb \pa\ipr A -\nn\\ 
&&\qquad\left. - \s c(x) c(x)-\a(1-\d)\f\cb(x)\cb(x) 
-\frac{1}{g^2}\s\f +\frac{1}{g} \g\cb(x) + \frac{1}{g}\sb c(x) \right]. 
\ee 
The constant Grassmann ghosts $\g$ and $\sb$ in \equ{effaction}  
implement the constraints  
\be\label{constraint} 
\int_\TT dx\ c(x) &=& 0\nn\\ 
\int_\TT dx\ \cb(x) &=& 0 
\ee 
which remove the (on a {\it finite} torus normalizable) constant 
modes of the ghost-fields in a translationally invariant fashion. 
In the framework of TQFT one can show that the partition function of 
the gauge-fixed quantum theory on a compact space-time manifold
otherwise would vanish due to these zero-modes of the
ghosts\cite{mi81,bs97,brs96}. The surface term $\frac{1}{\a g}\gb \pa\ipr A $
in\equ{effaction} vanishes on a finite torus with periodic boundary
conditions. The absence of this term on a torus
will eventually give rise to a non-trivial effective potential on the
moduli-space. We retain this surface term in order to manifestly
exhibit the BRST-symmetry of\equ{effaction}.  

The action\equ{effaction} is formally invariant under the action of
$\ts$ defined by: 
\ba{rclrcl} 
\ts A_\mu(x) &=& D^{A }_\mu c(x)&\ts c(x) &=& -\frac{g}{2}[ 
c (x), c (x)]-\frac{1}{g}\f\\ 
\ts \cb(x)&=& (\gb/g -\pa\ipr A(x))/\a -\d g[\cb(x), c(x)] &\ts\f&=&0\\ 
\ts \s &=& \sb & \ts \sb &=& [\f, \s]\\  
\ts \gb &=&\g & \ts \g &=& [\f, \gb]\\ 
\ts \p_i(x) &=& -g c(x)\p(x)_i & \ts \pb_i(x) &=& -g\pb_i(x) 
c(x) 
\ea{newbrs}
which is on-shell nilpotent only on the set of globally $SU(n)$ 
invariant functionals. 
Apart from an unorthodox transformation of the anti-ghost 
$\cb$ and the presence of additional constant ghosts, $\ts$ is similar 
to the conventional BRST-transformation, and defines 
the equivariant cohomology of physical observables. 

The constraints\equ{constraint} are
invariant with respect to \equ{newbrs} at points where 
\eq
\int_\TT dx \ts \cb(x)=0\qquad {\rm and}\qquad \int_\TT dx \ts c(x)=0
\eqn{sconstraint}
Dividing by the finite volume of the torus and taking the
thermodynamic limit, \equ{sconstraint} 
relates the global bosonic
ghosts $\gb,\f$ to vacuum expectation values of composite dynamical
fields 
\eq
\vev{\gb}=\a\d g^2 \vev{[\cb(x), c(x)]}\qquad{\rm and}\qquad \vev{\f}=
-g^2\vev{c^2(x)}
\eqn{relvev}
The relations\equ{relvev} show that the equivariant BRST-symmetry of
the model is unbroken in the thermodynamic limit only at very specific
values for the global ghosts. 
In what follows, we will see that the first of these relations
in\equ{relvev} is satisfied non-trivially in certain gauges and will explore
the consequences. 

The expectation value of an observable $\OO\in \Sigma$ is 
formally given by the path integral  
\eq 
\vev{\OO}=\NN \int d\f d\s d\sb d\gb d\g\  
\int\!\int [\DD \p\DD\pb\,\DD A\, \DD c\,\DD\cb]\ \OO\ \exp S_0,
\eqn{expect} 
which defines the perturbative loop expansion. We show in
Appendix~A that \equ{expect} is generally normalizable, i.e. that
$\vev{\one}\neq 0$ for an $SU(2)$ group. 
The global bosonic ghosts $\f,\s$ and $\gb$ introduced by the 
equivariant  BRST-algebra are  
moduli parameters of  the theory. We indicate in\equ{expect} that the  
integration over this finite dimensional moduli-space would usually  
be performed {\it after} the path integral over    
the dynamical fields. As far as the dynamical fields
are concerned, the moduli are parameters of the action.  
Note that the moduli space possesses a trivial BRST-symmetry.
 
With finite fermion-masses $m_i$, and/or anti-periodic boundary conditions 
for the fields  $\p_i$ and $\pb_i$, only the constant modes of the
connection $A_\m(x) $ are potentially troublesome in a
perturbative treatment of the theory. One can eliminate
these constant modes by requiring twisted boundary
conditions\cite{tH79}. These  boundary
conditions however break the global $SU(n)$ invariance of the gauge
theory explicitly\footnote{the
equivariant construction is then in fact not necessary, since twisted
boundary conditions also eliminate the constant
ghost modes} and  effectively
introduce external color flux\cite{tH79}. Although this is arguably the
most elegant treatment of the purely gluonic theory on a torus, twisted
boundary conditions require the absence of fields in the fundamental
representation. They also complicate the mode expansion and thus the
infinite volume limit, especially since the space of gauge orbits in
this case is disconnected. The global $SU(n)$ symmetry of the
equivariantly gauge fixed theory is, as we
shall see, {\it spontaneously} broken even  
for periodic boundary conditions at certain points in the parameter
space. The theory is thus seen to react extremely sensitively to any
external color flux. 

The constant  part $a_\m$ of a
field configuration  $A_\m (x)$ satisfying periodic boundary conditions   
\eq 
A_\m(x) =\frac{1}{g}a_\m +  \tilde A_\m(x) \qquad{\rm with } \qquad 
\int_\TT dx\ \tilde A_\m(x) =0  
\eqn{constA} 
can be treated as another moduli on which the 
perturbative expansion depends. Since  we
only evaluate correlation functions of local operators,
the integration over this moduli space is governed by the corresponding 
effective action. Due to translational
invariance of the constant modes, this effective action is
proportional to the volume of space-time. In the infinite volume
limit only the immediate vicinity of the absolute minimum (possibly degenerate)
 of the effective potential therefore contributes to the integration over 
the moduli-space. Contrary to supersymmetric gauge
theories, we do not obtain the complete dependence of the
effective potential on the moduli parameters. The assumption that
perturbation theory is a good asymptotic expansion in conjunction with
some physical criteria, however greatly constrains the minimum of the effective
potential. Only the minima which are stable fixed points can be
consistently incorporated in a perturbative expansion of the theory, since
moduli tend to  flow to such fixed points as the coupling $g$
becomes critical.  

From a physical point of
view, the minimum of the effective potential should occur for $a_\m=0$, since translational invariance implies that  
\eq 
\vev{A_\m(x)}=\frac{1}{g}\vev{a_\m}. 
\eqn{vevA} 
The discrete rotation group  of a symmetric torus would be
broken unless all 
$\vev{a_\m},\ \m=1,\dots,D$ are equal. Equivariant observables such as
$\tr F_\mn F_\rs$ or 
$\pb_i(x)\g_\m\p(x)$ furthermore would register  
a breakdown of  the  $SO(4)$-symmetry in the continuum 
limit for $\vev{A_\m}\neq 0$, unless this phase is gauge-equivalent
to one with vanishing $\vev{A_\m^\prime}=0$.  
A phase with  nontrivial  $\vev{A_\m}$ is therefore acceptable only if 
\eq 
\vev{a_\m} = U^\dagger(x)\pa_\m U(x)  
\eqn{Uvac} 
for some  $U(x)\in SU(n)$ on the torus. In this case the theory is
however physically equivalent to the one with $\vev{a_\m}=0$.  
Since the torus has only $1$-cycles, \equ{Uvac} implies
that $\vev{a_\m}$ is abelian, i.e.  
$[\vev{a_\m}, \vev{a_\n}]=0$. In Appendix~A we find these  abelian
Gribov copies of the trivial vacuum  
$A_\m=0$. They  are not removed by the equivariant construction on a
torus. The contribution to $\vev{A_\m}$ from these gauge copies
however cancel. Gauge invariant local   
physical observables do not depend on which of these copies is
selected, and we may  as well perturbatively
expand around the trivial  configuration $\vev{A_\m}=0$.
  
The global $SU(n)$ symmetry is also spontaneously broken by a non-vanishing
expectation value of any of the constant bosonic ghosts $\f,\s$ or  
$\gb$ of the equivariant gauge fixing. None of the local
functionals in the 
equivariant cohomology $\Sigma$ are order parameters for a  
spontaneous breakdown of the global $SU(n)$ symmetry.  We will
see, the the {\it perturbative} evaluation of
correlation functions in certain gauges is however sensitive to  whether and  
how the global $SU(n)$-symmetry is broken. The
situation is analogous to ordinary spontaneous symmetry breaking if 
one restricts the observables of the model to group invariants: the
perturbative calculation of  ``invariant'' correlations depends on
whether the vacuum breaks the symmetry or not, but none of these
invariant observables indicates which of the degenerate symmetry
breaking vacua has been selected. Furthermore, the Goldstone bosons of
the symmetry breaking {\it do not} lead to massless singularities in
correlations of the invariant ``observables''. For such a restricted
set of observables, 
the symmetry breaking is in a sense {\it hidden} and its effects
appear to have a {\it dynamical} origin.
   
Non-vanishing $\vev{\f}$ or $\vev{\s}$ 
implies that the ghost number is not conserved. The expectation value of
the bosonic ghost $\gb$ however conserves the ghost charge and 
 there is no a priori reason for it to vanish in the equivariant theory. 
   
\section{Fixed points of the gauge parameters} 

We begin by investigating the critical behavior of the dimensionless gauge
parameters  $\a$ and $\d$ for $g\rightarrow 0$. For $\d=0$, the gauge
parameter $\a$ was observed to flow to a non-trivial fixed
point\cite{ma78}. We will see that $\a$ flows to a fixed point
$\a_{\infty}$ for a wide range of values of the gauge parameter $\d$.
  
Multiplicative renormalization constants relating bare (B) to 
renormalized (R) dynamical fields and parameters are defined as usual 
\ba{rclrclrcl} 
Z_2^{1/2} (\p_R,\pb_R)&=&(\p_B, \pb_B)\  &Z_3^{1/2}  A_R &=&A_B \ 
&\tilde Z_3^{1/2} (c_R,\cb_R)&=&(c_B, \cb_B)\\ Z_g g_R&=&g_B&Z_\a 
\a_R &=&\a_B&Z_\d \d_R&=&\d_B 
\ea{rconst} 
At the one loop level,  $\d$ does not renormalize\cite{ba82}. 
Counter terms for the gluon polarization determine $Z_3$ as well as 
$Z_\a$.  Note that the usual Slavnov-Taylor identity for the longitudinal part
of the gluon propagator\cite{sl72} only requires 
equality of $Z_\a$ and $Z_3$ for $\d=0$. In generic covariant gauges
the equation of motion  of the Nakanishi-Lautrup field\equ{EMb} also involves 
the composite operator $[\cb(x), c(x)]$ and $Z_\a\neq Z_3$. To one
loop this is confirmed by an explicit calculation of the
renormalization constants  of\equ{rconst}. In the MS-scheme one
obtains in $D=4-2\e$ space-time dimensions 
\ba{rcl} 
Z_2&=&1-\frac{g^2 \m^{-2\e} (n^2-1)\a}{16\pi^2 n\e}+O(g^4)\\ 
&&\\ 
\tilde Z_3&=& 1+\frac{g^2 \m^{-2\e} n}{64\pi^2\e}(3-\a)+O(g^4)\\ 
&&\\ 
Z_3&=&1+\frac{g^2\m^{-2\e} 
}{16\pi^2\e}\left(\frac{13}{6}n-\frac{2}{3}n_f-\frac{\a n}{2}\right) 
+ O(g^4)\\ 
&&\\ 
Z_\a &=& 1+\frac{g^2\m^{-2\e} 
}{16\pi^2\e}\left(\frac{13}{6}n-\frac{2}{3}n_f-\frac{\a n}{2} 
+\d(1-\d)\a n\right) + O(g^4)\\  
&&\\  
Z_g&=&1-\frac{g^2 \m^{-2\e} }{32\pi^2\e}\b_0+ O(g^4) 
=1-\frac{g^2 \m^{-2\e}}{16\pi^2\e}\left(\frac{11}{6}n-\frac{1}{3}n_f\right)+ 
O(g^4)\\  
&&\\ 
Z_\d &=& 1+ O(g^4) 
\ea{zfac} 
(The fact that $Z_3=Z_\a$ also for $\d=1$ is not a consequence of the usual
Slavnov-Taylor identity\cite{de89}.)  Note that the  
renormalization constant $\tilde Z_3$  of the dynamical ghosts does
not depend on $\d$ at the one-loop level. In\equ{zfac} and most of the
following we dropped the index (R) for renormalized quantities.

For $\a=\a_\infty$ 
\eq 
\a_\infty(\d)=\frac{13 n- 4 n_f}{3 n (1-2\d+2\d^2)}    
\eqn{fixalpha} 
$Z_\a$ is independent of the renormalization scale $\m$ to order $g^2$. 
It is readily seen that $\a_\infty$ is the stable fixed point of the gauge 
parameter $\a$ which determines the asymptotic behavior of 
gauge dependent correlations.   
 
The relation\equ{rconst}  between bare and renormalized gauge
parameter  implies for the renormalized $\a$   
\eq 
\frac{{\rm d\,}\ln{\a}}{{\rm d\,}\ln\m} + \frac{{\rm 
d\,}\ln{Z_\a}}{{\rm d\,}\ln\m} =0  
\eqn{anomaldima} 
in any gauge other than Landau gauges. Integration of 
\equ{anomaldima} using the 1-loop expression for $Z_\a$  and 
\eq 
\frac{d 
g}{d\ln\m}=\b(g)=-\frac{g^3}{16\pi^2}\b_0+\dots\qquad{\rm with}\qquad 
\b_0=\frac{11}{3}n-\frac{2}{3}n_f  
\eqn{betafunction} 
shows that the gauge parameter $\a$  
approaches the fixed point $\a_\infty$ of\equ{fixalpha} for
sufficiently small coupling $g$ as 
\eq 
\a-\a_\infty=C\a g^{\frac{ 13 n -4 n_f}{11 n -2n_f }}\ 
(1+O(g^2)\ ) 
\eqn{runninga} 
where $C$ is an integration constant related to the choice of 
gauge at a particular (small) value of the coupling.  $C$ is related to $\a$
in a similar fashion as  $\L_{ASP}$ is related to the coupling $g$: it
is the RG-invariant ``gauge-parameter'' which governs the asymptotic flow 
of $\a$ as $\m\rightarrow \infty$.  Note that the critical exponent
in\equ{runninga} is independent of $\d$ and positive in the
asymptotically free regime when  
\eq 
n_f < \frac{13}{4} n < \frac{11}{2} n 
\eqn{rangenf} 
This implies that Landau gauge  is {\it not} the
asymptotically stable fixed point in the gauge  
parameter space if the number of families is not too large. Landau
gauge becomes the stable  
fixed point only when the exponent in\equ{runninga} is negative,
i.e. for $\frac{11}{2}n> n_f > \frac{13}{4}  
n$ flavors. In the case of QCD this occurs only for $5-8$~families.
Note also that $\a$ tends to  
$\a_\infty$ slower than $g^{13/11}$. The retained term of $Z_\a$ 
therefore dominates over the neglected terms of order $g^4$
in\equ{anomaldima} at sufficiently small  
coupling. For $n_f=n$,
$\a$ approaches $\a_\infty$ proportional to $g$.  We remark 
that the fixed point $\a_\infty$  as well as the critical exponent
in\equ{runninga} tend to  finite, non-vanishing values in the limit of
large $n$, a case which will interest us. In general,
\equ{runninga} describes a surface in the space of continuous
parameters $\a,\d,g,n_f/n$ of the model. We include the ratio $n_f/n$
as a parameter of the model, since asymptotic freedom for instance
depends on its value.   
 
For  fixed $n_f/n$, the perturbative expansion of an $SU(n)$ gauge theory is
selfconsistent in the vicinity of the curve in  
the coupling space $g,\a,\d$ corresponding to $g=0$, and
$\a=\a_\infty(\d)$ given by \equ{fixalpha}. The coupling constants $g$
and $\a$ can be traded for the RG-invariants $\L_{ASP}$ and $C$. 
RG-invariant quantities only depend on  $\L_{ASP}$, 
$C$ and $\d$,  and {\it physical} gauge 
invariant correlations should furthermore not depend on the 
last two.
An immediate consequence of this analysis is that the leading deep Euclidean 
behavior of the longitudinal gluon propagator is effectively described 
by the tree-level propagator in the gauge $\a_\infty(\d)$: corrections to 
this asymptotic behavior vanish logarithmically for sufficiently 
large momentum transfers.  At the fixed point $\a_\infty$ the term $\pa/\pa\a$
in the RG-equation only induces corrections that are
{\it analytic} in $g$ and which vanish for $g\rightarrow 0$.

It is perhaps surprising that a gauge parameter such as $\a$ 
depends on the renormalization scale at all. $\a$ was introduced
as a coupling constant of the TQFT on the gauge group\cite{bs97} and the
scaling-functions  
of a TQFT should vanish. The coupling constants of a TQFT are therefore 
not expected to depend on a renormalization scale.  We 
do not have an exhaustive explanation for this apparent
paradox. However, if one  distinguishes 
between the definition of the TQFT and the {\it
perturbative} calculation giving the apparent scale dependence of $\a$,
this may not be so paradoxical.  We wish to point out that only {\it
one} Gribov copy of a gauge field configuration 
contributes in the perturbative 
evaluation of a Green's function (the one in the vicinity of $A_\m(x)=0$
). On the other hand, the conclusion 
that a parameter of the TQFT does not depend continuously on a scale
requires the contributions from all stationary points, 
that is from {\it all} Gribov copies of a configuration. It is
therefore conceivable that the asymptotic behavior of the 
longitudinal gluon propagator evaluated within a {\it single} 
Gribov region depends on the renormalization scale,  whereas it would be scale
independent, if the contribution from all other Gribov copies were
considered as well (it is known that the ``size'' of a particular  Gribov
region is scale dependent\cite{gr78}). Perturbatively one verifies only
that gauge-dependent quantities evaluated in a {\it single} 
Gribov region, asymptotically require an adjustment of the gauge
parameter with the scale $\mu$. At the fixed point $\a_\infty$ this
scale dependence vanishes and the contribution from
other Gribov copies to the asymptotic behavior of gauge dependent
Green's functions can be ignored. We find the existence of such a
finite fixed point $\a_\infty$ remarkable.

The gauge invariant correlation functions one is ultimately interested
in should not depend on whether all gauge copies of a configuration
are taken into account or only one. Since the ``number'' of such
copies does not change within a connected sector of the orbit
space\cite{bs97}, the difference between the two evaluations of gauge
independent Green's functions within a single topological sector can
be absorbed in the normalization
of the path-integral. The orbit space on a torus 
is connected, and one should therefore obtain the
 asymptotic expansion of physical correlation functions by perturbing around a
{\it single} classical solution satisfying the gauge
condition. (See\cite{brs96} for a discussion of the dependence of the
normalization on the sector for an $SU(2)$ gauge theory on $S_4$.)  

We have found the fixed points in the parameter space.
In the next section we consider fixed points of the moduli space. We
will see that perturbation theory is 
consistent in a phase where the global
$SU(n)$ symmetry is spontaneously broken to $U^{n-1}$ for particular 
values of the gauge parameter $\d$.  Although this consistency
requirement for the {\it perturbative} expansion is gauge dependent,
the fact that the global $SU(n)$ 
symmetry is broken to $U^{n-1}$ perhaps is not.  
We will find that our moduli space is however consistent with such a
symmetry breaking only on a subset of  
the fixed points in the parameter space. For other values of the
parameters, non-perturbative configurations  are probably 
relevant even asymptotically -- just as contributions from  
other Gribov regions probably would stabilize $\a$ in
gauges $\a\neq \a_\infty$. At particular points in the parameter
space, perturbation theory indicates that the moduli space might be
adequate to describe this symmetry breakdown and that other
non-perturbative phenomena could be of sub-leading importance.

\section{Fixed points of the moduli space}  
Perturbation theory is not entirely determined by the UV-fixed points
of the parameters.
It also depends on the phase,  which usually is encoded in the
expectation values of scalar fields. In the case of gauge theories
these are composite 
operators, whose expectation values generally will not be described by
the moduli of our model. We can however perturbatively examine whether
this restricted  moduli 
space allows for non-trivial expectation values at
{\it particular} values of the model parameters. 

Perturbation
theory  on the torus in covariant gauges 
depends on the moduli-space $\MM$  
\eq 
\MM=\{\gb,\f,\s, a_\m\} 
\eqn{modulispace} 
The measure on
this moduli-space is induced by integrating out the  dynamical
fields, which gives the effective
potential for the moduli.
In TQFT's the semi-classical  measure on the 
moduli-space is {\it exact},  and in supersymmetric theories it is
also severely constrained by holomorphicity\cite{se94}. 

In contrast to supersymmetric gauge theories, this measure on the
moduli space will not be obtained exactly for the 
gauge theories that interest us. In a translationally 
invariant model, the effective action on the moduli space is however
proportional to  the space-time volume of the manifold.  In the
thermodynamic limit the integration over the
moduli-space is thus effectively 
restricted  to regions arbitrary close to absolute  
minima of the effective potential. To
evaluate the integral over the moduli space in the thermodynamic limit
in four space-time dimensions, we in principle only need to know the
scale of these minima and the behavior of the effective potential in
their immediate vicinity. If a compact 
group relates degenerate absolute minima of the potential and the
observables are invariants under this group, it furthermore  suffices to 
evaluate the moduli at {\it one} of the degenerate
absolute minima only. The group is then  spontaneously
broken to the subgroup which leaves this minimum of the effective
potential invariant.  

Suppose for a moment that the effective potential of the
asymptotically free theory in four dimensions has a unique
RG-invariant non-trivial absolute minimum modulo group invariance.
In the absence of dimensionful parameters, the scale $\k$ associated with this minimum\footnote{The canonical 
dimension of  $\f$ in \equ{modulispace} vanishes, but the effective
potential conserves ghost number (although 
its minima need not to), and is thus a function of $\s\f$ rather
than $\f$ itself.}  is a function of the renormalization point $\m$
and the dimensionless 
coupling constants $g,\a,\d$ only. Since the minimum is a fixed point of
the moduli space and the theory is asymptotically free, $\k$ must be
proportional to $\L_{ASP}$ and its 
dependence on $\m$ and $g$ at weak coupling therefore is 
\eq 
\ln{\frac{\k^2}{4 \pi\m^2}}= - \frac{16\pi^2}{\b_0g^2} 
-\frac{\b_1}{\b_0^2}\ln\frac{\b_0 g^2}{16 \pi^2} + c_\k + O(g^2) 
\eqn{scale}  
where 
\ba{rcl}  
\b_0&=&\frac{11}{3}n-\frac{2}{3}n_f\\ 
&&\\ 
\b_1&=&\frac{34}{3}n^2-\frac{13}{3}n n_f +\frac{n_f}{n} 
\ea{bcoeffs} 
are the first two coefficients of the $\b$-function.
Since the leading coefficient of $1/g^2$ in\equ{scale} cannot
vanish, $\k$ has to be the scale of the {\it one-loop} effective
potential also. In fact, $\k$ has to be  associated with a
non-trivial  {\it absolute  minimum} of the one-loop potential in the
case we are considering, since
the RG-invariant absolute minimum of the effective
potential defines a mass scale. Conversely, a non-trivial
absolute minimum of the one-loop effective potential which is RG-invariant and 
unique modulo group invariance {\it defines} an asymptotic mass scale
which can be 
related to other definitions of $\L_{ASP}$ by a two-loop
calculation\cite{ce79}. It should be emphasized that an RG-invariant minimum of the effective
potential on the moduli space gives an  {\it  asymptotic} scale, whose
coefficients in the expansion\equ{scale} can be reliably calculated {\it perturbatively}. 

The coefficient of the $g^2$ term in\equ{scale} is determined by the
one-loop  effective potential and {\it must} be 
$-(4\pi)^2/\b_0$ at a fixed 
point of the moduli space.  Depending on whether $\L_{ASP}\gg\k$ or
$\L_{ASP}\ll\k$ at weak coupling, one otherwise concludes that the
only fixed point of the moduli space is the trivial one with
$\k/\L_{ASP}=0$ or the one at infinity. In the latter case the
perturbative evaluation of the potential is certainly
inconsistent, since it gives rise to an arbitrary large mass
scale\footnote{the usual $\L_{ASP}$ should set the  mass
  scale and the asymptotic expansion is not expected to  depend on masses much
  larger than $\L_{ASP}$.}. It should be emphasized that this first 
coefficient of the expansion\equ{scale} is entirely determined by the
classical term of the effective potential and its {\it divergent}
one-loop contribution. Since the effective potential is only
UV-divergent, a perturbative calculation of this coefficient is
reliable in asymptotically free gauge theories. 

Whether $\k$ 
is truly marginal for $g\rightarrow 0$ is  determined by the coefficient 
of the $\ln g^2$ term in \equ{scale}. It is given by the
one-loop  anomalous dimension of the 
order parameter. Since the effective potential is itself
RG-invariant, the coefficients of the $1/g^2$- and $\ln g^2$-terms
in fact are related by the RG-equation: near a fixed point in the
parameter space, $\k$ is of the
form\equ{scale} only if the 
critical exponent of the moduli parameter vanishes.  The relation
between the first two coefficients in\equ{scale} also greatly
facilitates the search for fixed points of the whole moduli 
space, since one merely needs to compute the critical exponents of the
moduli on the curve\equ{fixalpha} in the gauge parameter space. 
As shown in Appendix~B, the anomalous dimension of a moduli is
also determined by  the divergence of the one-loop effective potential
and the associated classical counterterm.

A two-loop calculation of
the effective potential would in addition uniquely determine the
coefficient $c_\k$ in 
\equ{scale} and thus relate $\k$ to any other asymptotic scale
parameter quantitatively\cite{ce79}.  To qualitatively analyze the
moduli space,  
the relation of $\k$ to other asymptotic scale
parameters is not required and will not be obtained here.  We
nevertheless emphasize, that a unique
RG-invariant non-trivial absolute minimum of the effective potential
on the moduli space is {\it perturbatively} related to other
definitions of the asymptotic scale parameter in ordinary gauge
theories.

It might appear that the converse is not necessarily true, and
that a non-trivial absolute minimum of the one- and two-loop effective
potential with the above properties does not necessarily imply the
same for the full effective potential. The
argument above however shows that this would question the {\it
  validity} of perturbative results in ordinary gauge theories  since
non-perturbative effects would have to redefine the perturbative
 asymptotic scale\footnote{In $D=2$ space-time
dimensions the order parameter is dimensionless and the effective
potential {\it does not} define an asymptotic scale and there is no
spontaneous symmetry breaking\cite{co73b}. In
supersymmetric gauge theories in four dimensions, non-perturbative
corrections are asymptotically important\cite{sh86}  and their contribution is
required to unambiguously define an asymptotic scale. In theories which
are not asymptotically free, such as $\phi^4$, our argument obviously
does not apply.}. Perturbation
theory by  itself ought to be  
adequate in determining critical exponents
etc. of ordinary gauge theories, and the perturbative asymptotic
scale can be given a physical interpretation in this case.  This
implies that the 
RG-invariant non-trivial absolute minimum of the two-loop  effective
potential which  defines an {\it asymptotic} scale
cannot simply be absent from the full effective potential. This
argument for the 
existence of such a scale-defining absolute minimum also in the full
effective potential does {\it not} mean that
the two-loop effective potential is {\it exact} in ordinary gauge
theories. It only implies that a RG-invariant scale
set by the absolute minimum of the effective potential can be used to normalize
the model also perturbatively and that the full effective
potential must have a non-trivial absolute minimum, if the one-loop potential
does and if this minimum is RG invariant. In the following we
therefore only verify whether the one-loop 
effective potential on the moduli space has non-trivial RG-invariant
absolute minima at fixed points of the parameter space.
 
Let us first assume that ghost number and angular momentum
are conserved, i.e. that
$\vev{\f}=\vev{\s}=\vev{a_\m}=0$. In this case we need to  consider only the
dependence of the effective potential on the moduli
$\gb$. Generalizing the calculation for an $SU(2)$ gauge group in\cite{bs97} to
$SU(n)$, we obtain in Appendix~B  the
dependence of the one-loop effective potential on the eigenvalues $\lm_i, 
i=1,\dots,n$ of the traceless $n\times n$ hermitian matrix $i\gb$  
\eq 
V_{1-loop}=\sum_{1\leq i<j\leq n}
\frac{(\lm_i-\lm_j)^2\d^2}{32\pi^2}\ln\left[\frac{(\lm_i-\lm_j)^2\d^2}{e\k^4}\right]  
\eqn{veff} 
Note that the full effective potential also depends only on 
eigenvalue differences, and that the logarithm in\equ{veff} is
typical for the one-loop approximation.
Crucial for the following is that\equ{veff} has a unique and
non-trivial {\it absolute
minimum} (modulo $SU(n)$) and the dependence of  
the associated scale $\k$ in\equ{veff} on the renormalization point 
$\m$ and the coupling constants. To this order in the loop 
expansion we obtain in Appendix~B that,  
\eq 
\ln \frac{\k^2}{4\pi\mu^2}=-\frac{(4\pi)^2}{n\d^2\a 
g^2}+1-\g_E +O(\ln g^2, g^2)   
\eqn{scalekap}
 in the MS-scheme. 
Comparing with\equ{scale} we see that $\k$ is an asymptotic scale
only in gauges where  
\eq 
n\d^2\a\ {\raise-.7ex\hbox{$\stackrel{{\displaystyle 
\longrightarrow}}{_{g\rightarrow 0}}$}}\ 
\b_0    
\eqn{marginal} 
The intersections of this curve with $\a_\infty(\d)$ of 
\equ{fixalpha} occur for gauge 
parameters $\d=\d^{\pm}_\infty$  
\eq 
\d^{\pm}_\infty= \frac{11n-2n_f \pm\sqrt{2(n-n_f)(11 n-2 n_f)}}{9 n} 
\eqn{deltafix}
For a given value of $n_f/n$, the two points  in the gauge parameter
space
$(\a_\infty(\d^+_\infty), \d^+_\infty)$ and $(\a_\infty(\d^-_\infty),
\d^-_\infty)$,  where the scale $\k$ is comparable to $\L_{ASP}$ on the curve 
$\a=\a_\infty(\d)$ define critical covariant gauges
(CCG).  Remarkably, we only find real intersections 
$\a_\infty(\d^{\pm}_\infty),\d^{\pm}_\infty$ in the asymptotically
free regime when 
the number of quark flavors does not exceed the number of colors   
\eq 
n_f\leq n 
\eqn{confphase} 
Note that for $n_f>n$, $n\d^2\a_\infty<\b_0$ for any real value of 
$\d$. The scale $\k$ in\equ{scalekap} is therefore irrelevant 
compared to $\L_{ASP}$ for ${11\over 2}n>n_f>n$. In the space of parameters
$\d,\a,n_f/n$ the condition\equ{marginal} is valid on a surface
which intersects the surface where\equ{fixalpha} is
satisfied on a critical curve. Projections of this curve onto the
$(\d,\a)$- and $(n_f/n,\d)$-planes are shown in Fig.~1. Only on this
critical curve in the parameter space, $\k$ and $\L_{ASP}$ are {\it
both} asymptotically relevant scales and proportional to each
other. The perturbatively  
inaccessible region is shaded grey in Fig.~1 and given by
$n\a_\infty\d^2> \b_0$, since $\k\gg\L_{ASP}$ in this case.
We emphasize that\equ{confphase} and\equ{deltafix} are consistency
requirements  for a {\it perturbative} evaluation. In other gauges (or
when $n_f>n$) there is no non-trivial fixed point of the moduli $\gb$.  In
the grey region of Fig.~1 perturbation theory is not consistent
and non-perturbative field configurations are perhaps 
important even asymptotically. In the white region of
Fig.~1, the scale $\k$ is irrelevant and the only fixed point of our
moduli-space is the trivial one.  The perturbatively interesting
region is the phase boundary where $\k\sim\L_{ASP}$.
\vskip -0.2in
\epsfbox{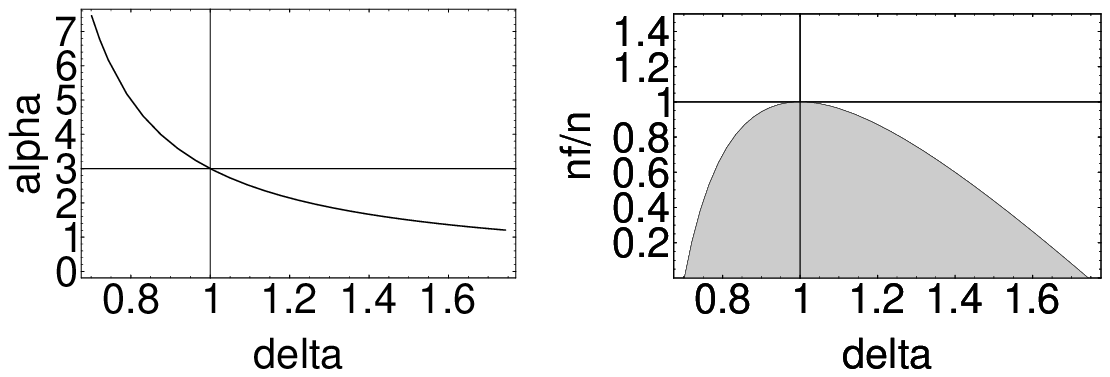}
\vskip -0.9in {\small\baselineskip 5pt 
\noindent Fig.~1: Projection of the critical
curve onto the ($\d,\a$)- and ($\d, n_f/n$)- planes. The two points on
this critical curve for a fixed value
of $n_f/n<1$ define CCG. The perturbatively
inaccessible region in the parameter space is shaded grey in the ($\d,
n_f/n$)-plane. The origin of the axes is at the critical point
($n_f=n,\d=1,\a=3$). }   

The gauge group of QCD is $SU(3)$ and there are  $n_f=6$ quark
flavors. Since $n_f=6>3=n$ in this case, the  perturbative analysis indicates
that $\k$ is an irrelevant asymptotic  scale for QCD. However, only $2-3$
light quark flavors seem to be important for the low-energy
dynamics of QCD 
and heavy quark flavors are furthermore unstable when flavor-changing weak 
interactions are included\footnote{The 
coupling to photons invalidates our perturbative analysis, which
depends heavily on asymptotic freedom}. It is not
inconsistent phenomenologically that $\k$ 
is irrelevant at energies much above the $J/\Psi$ 
threshold where more than $3$ quark flavors are dynamical and where
ordinary perturbation theory works reasonably well.  

To qualitatively describe the dynamics of QCD below a few $GeV$, an 
$SU(3)$ gauge theory with only  $3$ or even less dynamical quark flavors 
appears to be a reasonable model, especially when weak
interactions are ignored.  Heavy quark flavours appear to be
adequately described by heavy quark expansions\cite{ag94} around $\m_f\sim
\infty$. The purely gluonic theory with $n_f=0$
and/or the large $n$ limit of $SU(n)$ gauge theories are furthermore of
theoretical interest, since Wilson's 
criterium\cite{wi74} for absolute confinement is only valid in the absence
of dynamical quarks. In these limiting cases, $\k$ {\it is} an
asymptotic scale in CCG. 

Let us therefore explicitely verify that $\vev{\gb}$ is a truly marginal
parameter in a (hypothetical) $SU(n)$ gauge theory with $n_f\leq n$
quark flavors in CCG. The
critical exponent of $\gb$ vanishes if the anomalous dimension of
$\gb^2$ is of order $g^4$. The renormalization constant 
$Z_{\gb^2}$ relating bare and renormalized moduli, $\gb^2_R 
Z_{\gb^2}=\gb^2_B$, is calculated in Appendix~B to order $g^2$. For 
arbitrary gauge parameters $\d,\a$ one obtains
\eq 
Z_{\gb^2}=1-\frac{g^2\m^{-2\e}n}{32\pi^2\e}\left(3+\a (1-2\d)\right)
+O(g^4) \eqn{renormgb}

In CCG defined by the fixed points\equ{fixalpha} and \equ{deltafix}, 
the coefficient of $g^2$ in\equ{renormgb} vanishes 
and the critical exponent of $\gb^2$ thus vanishes as well. We 
thus have explicitly verified the relation between the two coefficients in
\equ{scale} at fixed points of the parameter space.
The non-trivial absolute minima of the effective potential defining the scale
$\k$ therefore do not depend on the value of the coupling
$g$ and define an asymptotic RG-invariant scale $\k$ which is
a physical mass scale in CCG\footnote{Note that for the  
extreme case $n_f=n$, the two fixed points coalesce at
$\a_\infty=3,\d^\pm_\infty=1$, a kind of Yennie-gauge\cite{ye61} with  
nice infrared behavior of the quark propagator\cite{it80}, and
vanishing critical exponents of the  
ghost and gluon fields }. As far as the perturbative evaluation of
correlation functions invariant under the global $SU(n)$ symmetry in
CCG is concerned, $\gb$ in\equ{effaction} can be diagonalized and its 
eigenvalues treated as mass-parameters proportional to
$\L^2_{ASP}$ that do not depend on the coupling. 

As remarked earlier, the absolute minimum of the one-loop
effective potential\equ{veff} is unique modulo $SU(n)$
transformations. It is attained when none of the eigenvalue
differences vanish, and the $SU(n)$ group is therefore maximally broken to
$U^{n-1}$. Since the difference of two eigenvalues is proportional to
the asymptotic scale parameter and furthermore RG-invariant, a variant
of our previous argument indicates that the absolute minimum of the full
effective potential also only possesses a $U^{n-1}$ invariance in CCG.
Higher order- and non-perturbative- corrections could only lift
degeneracies among the eigenvalues and further break the symmetry. The
difference  of any two eigenvalues (for instance the lowest two) at
the fixed point of the moduli space however defines a physical and
asymptotic scale in CCG which cannot vanish in the full effective
potential either. 

We next verify our previous assumption that ghost number is conserved, 
i.e. that a non-trivial fixed point with $\vev{\f\s}\neq 0$  would be {\it 
perturbatively} inconsistent. We only need to show that the critical 
exponent of  $\s\f$ does not vanish anywhere on the curve 
$\a_\infty(\d)$ of\equ{fixalpha} for $n_f$ in  the domain\equ{rangenf}.   
A one-loop calculation similar to the previous one gives for the
renormalization constant $Z_{\s\f}$  
relating  the bare and renormalized fields
\eq 
Z_{\s\f}=1+\frac{n g^2\m^{-2\e}}{32\pi^2\e}\left(\a\d(1-2\d) -3\right) 
+ O(g^4)  
\eqn{zsigphi} 
The critical exponent of $\vev{\s\f}$  therefore vanishes only if
$\a_\infty\d(1-2\d)=3$.  Demanding also \equ{fixalpha},
this occurs only  
when the gauge parameter $\d$ is a solution of the quadratic equation   
\eq 
\d(1-2\d)(13 n -4 n_f) = 9 n (1-2\d+2\d^2) 
\eqn{detdelta} 
\equ{detdelta} which has real solutions only for 
\eq 
n_f>\frac{-5+18\sqrt {2}}{4} n >\frac{13}{4} n 
\eqn{bound}
But $n_f>\frac{13}{4} n$ is outside the range\equ{rangenf} for which
$\a$ approaches  the fixed point $\a_\infty(\d)$
asymptotically. Furthermore \equ{zsigphi} is not of order
$g^4$ at the fixed point $\a=0$ for  
$n_f>\frac {13}{4}n$ either. The assumption that ghost number is
broken therefore would be perturbatively inconsistent.

Consider finally the moduli $a_\m$, the constant parts of
the gauge field. For a non-trivial fixed point with an anomalous
dimension of order $g^4$ the term of order $g^2$ of the
renormalization constant $Z_{a}=Z_3
Z_g^2$ would have to vanish. From \equ{zfac} we see that this only
occurs for $\a=-3$ -- clearly an impossible value for the gauge
parameter. When $\a>0$, the anomalous dimension of $a_\m$ is 
negative, and the only perturbatively stable fixed
point is $a_\m=0$.
   
To summarize: the perturbative  analysis of this section indicates that 
the global $SU(n)$-symmetry is spontaneously broken to 
$U(1)^{n-1}$ for $n_f\leq n$
quark flavors in CCG. Ghost number on the
other hand appears to be conserved. The
non-trivial expectation value $\vev{i\d\gb}$ is  characterized by a
RG-invariant scale $\k$ at two critical points of the parameter space
\eq 
(g=0,\a=\a_\infty(\d^{\pm}_\infty), \d=\d^{\pm}_\infty)
\eqn{fixpoint}  given by\equ{fixalpha},
and\equ{deltafix}. We next
investigate qualitative implications for the low
energy dynamics of the theory from the spontaneous breakdown of the
global $SU(n)$ symmetry in  CCG.  

\section{The broken phase} Since the 
equivariantly gauge fixed $SU(n)$ theory is {\it covariant} and
translationally invariant in the infinite volume limit, 
spontaneously broken symmetries should give us access to the
non-perturbative infrared dynamics of the model by  Goldstone's
theorem\cite{go61}. 

A non-trivial expectation value  $\vev{\gb}\neq0 $ {\it
does  not} imply a
breakdown of the equivariant BRST-symmetry\footnote{The question of a
global anomaly in the BRST-symmetry of the equivariantly quantized
continuum model for certain base manifolds was addressed
in\cite{bs97}.}. $\vev{\ts \cb(x)}$ in fact is the
conjugate variable to $\gb$ in\equ{effaction} in the sense that
\eq
\pad{V_{eff} (\gb)}{\gb^a} =\vev{ \ts \cb^a(x)/g}_\gb =
\vev{\gb^a/(\a g^2)- \d f^{abc} \cb^b(x) c^c(x)}_\gb
\eqn{EMgb}
where we have used the equation of motion for $\gb$ and translational
invariance. A {\it
necessary} condition for an unbroken equivariant BRST-symmetry
therefore is that
$\gb$ is  an extremum of $V_{eff}$. Conversely, if we require an
unbroken BRST-symmetry\equ{newbrs}, the possible expectation values for the
moduli $\gb$ are solutions to the {\it gap equation} 
\eq
\vev{g^2\d\a f^{abc} \cb^b(x) c^c(x)}_\gb =\gb^a
\eqn{gap}
Note that\equ{gap}  in general is a somewhat weaker statement 
than that $\vev{\gb}$ correspond to an {\it absolute minimum} of the
effective potential. In the last section we saw that non-trivial
RG-invariant solutions to\equ{gap} exist in CCG.

An superficially similar equation to\equ{gap} has previously been
obtained in the thermodynamic limit of lattice gauge theory in Landau
gauge as a necessary {\it horizon condition} for effectively constraining
the gauge theory to the fundamental modular
region\cite{zw94}. Apart from the fact that\equ{gap} also introduces a
mass scale via a consistency requirement, this gap equation differs
from the {\it horizon condition} in many respects. Perhaps the most
obvious difference is that the Landau-type gauges 
of\cite{zw94} cannot be investigated in our way, since they do
not correspond to perturbatively stable fixed points of the gluonic
theory. Furthermore, a non-trivial solution 
to\equ{gap} breaks the global $SU(n)$ symmetry, which was assumed not
to be broken in the lattice gauge theory\cite{zw94}. Finally, it was
shown in\cite{ms94} that a non-trivial solution of
the {\it horizon condition} spontaneously breaks the BRST-symmetry of the
lattice gauge theory in the thermodynamic limit.
The {\it gap equation}\equ{gap} in our case on the other hand is a
 minimal requirement for the equivariant BRST-symmetry of the model to
 remain unbroken.

Although\equ{gap} is only a necessary condition, we assume in the
following that the equivariant BRST-symmetry is not broken in CCG.  
Perturbation theory gives no indication to the contrary and the
breaking of BRST found in\cite{bs97} for certain base manifolds does not
occur for a torus. It should however be noted that the naive
BRST-symmetry of covariant gauges\cite{it80} {\it is} broken -- it
would give a vanishing partition function and was the reason for
restricting to an equivariant cohomology. It is therefore
perhaps not so surprising that we find non-trivial fixed points of the
moduli space with this equivariant construction in certain gauges.

\equ{gap} relates $\vev{\gb}$ to the expectation value of a
{\it local} composite operator. To find the Goldstone bosons of the
spontaneously broken symmetry, we first break the $SU(n)$
symmetry explicitly and extend the action\equ{effaction} by a
BRST-invariant local source term
\eq
S_0 \rightarrow S[\r]=S_0 +2 \a \int \tr \rho(x) \ts \cb(x)
\eqn{breaking}
We now study local Ward-identities of this extended model in the limit
of vanishing almost constant source $\r(x)$ in the infinite
volume limit. 

To derive the relevant Ward-identity, consider the following local
infinitesimal variation of the fields in the functional integral
\ba{rclrcl}
\d A_\m(x) &=& D_\m^A \t(x) = \pa_\m\t(x) + g[A_\m(x),\,\t(x)] &&&\\
\d \p_i(x) &=& -g\t(x)\p_i(x) &\d \pb_i(x) &=& g\t(x)\pb_i(x) \\
\d c(x) &=& g[c(x),\,\t(x)] &\d \cb(x) &=& g[\cb(x),\,\t(x)] \\
\d \gb &=&\d\g=\d \sb =\d\s =\d \f = 0\\
\ea{variation}
The transformation of the fermion- and gauge-fields in\equ{variation}
is just an infinitesimal gauge transformation and thus only the gauge
fixing part of the action\equ{breaking} changes to first order in
$\t(x)$ 
\ba{rcl}
\fud{S[\r]}{\t^a(x)} &=& -\pa_\m (J_\m^a (x) + D^{ab}_\m \r^b(x)) +
 f^{abc}((\pa_\m A^b_\m(x)) + \a\d g^2 f^{bde} \cb^d(x) c^e(x)) \r^c(x) - \\
&&\qquad - f^{abc} (\g^b\cb^c(x) -g\d\gb^b f^{cde} \cb^d(x) c^e(x) +\sb^b c^c(x) -\\
&&\qquad - \frac{g}{2}\s^b f^{cde}c^d(x) c^e(x) -\frac{g\a (1-\d)}{2}\f^b f^{cde}\cb^d(x) \cb^e(x)\,
)
\ea{varS}
The variation\equ{variation} gives the color current 
\ba{rcl}
J_\m^a(x) &=& -\ts D_\m^{ab}\cb^b(x) -\a^{-1} f^{abc}\gb^b A^c_\m(x) =\\ 
          &=&\a^{-1} D_\m^{ab} (\pa\ipr A^b(x) ) +g\d f^{abc} c^b(x)
D_\m^{cd} \cb^d(x) -g (1-\d) f^{abc} \cb^b(x) D_\m^{cd} c^d(x)
\ea{colcurr}
without use of the equations of motion.
The current\equ{colcurr} does not explicitly depend  on the moduli and
in fact is just the usual one for
gauges with quartic ghost interaction\cite{ba82}. Contrary to 
the standard case however, this current is {\it not} $\ts$-exact for $\gb\neq
0$. Note that the constant 
ghosts do not transform under the {\it local} variation\equ{variation} and
that the change of the action is therefore {\it not} proportional to
the divergence of the color current\equ{colcurr}. This somewhat 
complicates the Ward identities, but we will see that
the additional terms are irrelevant at vanishing momentum transfer.
Because the gauge symmetry is not
anomalous, the measure of the path integral is invariant under the change of
variables\equ{variation}. We thus have the Ward-identity
\eq
\vev{\fud{\OO}{\t^a(x)}}_\r
+\vev{\OO \fud{S[\r]}{\t^a(x)} }_\r = 0
\eqn{wident}
where $\vev{\dots}_\r$ indicates that the expectation value is for
fixed  source $\r(x)$ and $\OO$ is some functional of the fields. With
$\OO =\OO^a(x)$
\eq
 \OO^a(x)=g\a\d f^{abc} \cb^b(x) c^c(x)
\eqn{Odef}
\equ{wident} at vanishing momentum transfer in the limit of
infinite space-time volume becomes
\ba{l} 
\int d^4y\ \vev{\OO^a(x) \pa\ipr (J^b(y)+D^{bc}\r^c(y))\ }_\r= \vev{g
f^{acb} \OO^c(x)}_\r +\\
\qquad +g\int d^4y f^{bcd}\r^c(y)\fud{}{\r^d(y)}\vev{\OO^a(x)}_\r \\
\qquad +\vev{\OO^a(x)\ gf^{bcd}\left(\gb^c\pad{}{\gb^d} + \g^c\pad{}{\g^d} +
\sb^c\pad{}{\sb^d}+\s^c\pad{}{\s^d} +\f^c\pad{}{\f^d}\right) S[\r]}_\r
\ea{brokenwi}
Since $\OO^a(x)$ does not explicitly depend on constant ghosts,
the last term in\equ{brokenwi} is proportional to equations of motion
of the moduli and vanishes.  In  the limit
$\rho(x)\rightarrow 0$ of a
slowly varying source, \equ{brokenwi}
 requires that 
\eq
\int d^4x\ \vev{\pa\ipr J^a(x)\ \OO^b(y)}=\int d^4x\ \vev{\pa\ipr
J^a(x)\ g\a\d f^{bcd}\cb^c(y) c^d(y)}= f^{abc}\gb^c
\eqn{Goldstones}
where we used\equ{EMgb} to express the expectation value
of the operator $\vev{\OO^a(x)}_{\r\rightarrow 0} $ in terms of $\gb^a$. 
Since the RHS of\equ{Goldstones} does not vanish at the non-trivial minimum
of the effective potential in CCG, \equ{Goldstones}
implies a long range behavior of the correlation function 
\eq
\vev{J_\m^a(x)\ \OO^b(0)}\ {\raise-.7ex\hbox{$\stackrel{{\displaystyle   
\longrightarrow}}{_{x^2\rightarrow \infty}}$}}\
 f^{abc}\gb^c{x_\m\over 2\pi^2 x^4} 
\eqn{goldrange} 
which is the signature of massless Goldstone modes. Note that
the last step assumes translational {\it and}
Euclidean $SO(4)$ invariance of the thermodynamic limit. 

The relation \equ{goldrange} between $\vev{\gb}$ and the correlation
function of the composite operator $\OO^a(x)$ of\equ{Odef} with the
color current\equ{colcurr} does not appear to be RG-invariant, since
$\OO^a(x)$ generally mixes with 
$\pa\ipr A$. It is however straightforward to check that the
divergences on the LHS of\equ{goldrange} cancel to order $g^2$. We
have observed earlier that $Z_{\gb^2}=1+O(g^4)$ {\it
precisely} at the fixed points\equ{fixpoint} defining CCG. The
relation\equ{goldrange} is therefore consistent in the weak
coupling regime of CCG.

However, divergences in the correlation function\equ{goldrange} only
cancel to first order in the loop expansion for the {\it combination}
of composite operators in\equ{colcurr}. The mixing between  individual 
terms in $J^a$  implies that the
Goldstone pole must also be present in every correlation
function of $\OO^a(x)$ with any composite operator appearing in the
color current $J^a$ and in particular in the 1PR part of $J^a$
\eq
\vev{\pa_\m \pa\ipr A^a(x)\ \OO^b(0)}\
{\raise-.7ex\hbox{$\stackrel{{\displaystyle    
\longrightarrow}}{_{x^2\rightarrow \infty}}$}}\
 -{2 x_\m Z(\m)\over x^4} f^{abc}\gb^c
\eqn{GoldstoneA} 
The residue $Z(\m)$ in\equ{GoldstoneA}  depends on the
renormalization point $\m$ even at arbitrary weak coupling -- only the
residue in\equ{goldrange}  of the Goldstone contribution to the
correlation of $\OO^a$  with
the full current $J_\m^a$ is RG-invariant in CCG in the limit
$\m\rightarrow\infty$. $Z(\m)$ however cannot vanish for arbitrary values
of the renormalization scale, because the operator $\pa_\m 
\pa\ipr A^a$ is necessary for a cancelation of the one-loop divergences on the
LHS of\equ{goldrange}. A coupling to the Goldstone
modes is therefore inevitably generated in\equ{GoldstoneA} by a
change of the renormalization scale. 

In the translationally invariant theory, \equ{GoldstoneA} requires a
long range behavior of the ``mixed'' propagator of the form
\eq
\vev{\pa\ipr A^a(x)\ \OO^b(0)}\ {\raise-.7ex\hbox{$\stackrel{{\displaystyle   
\longrightarrow}}{_{x^2\rightarrow \infty}}$}}\
 {Z(\m)\over x^2} f^{abc}\gb^c
\eqn{GoldstoneDA} 

We now argue that a long-range behavior of the longitudinal
gluon propagator 
\eq
\vev{\pa\ipr A^a(x)\ \pa\ipr
A^b(0)}{\raise-.7ex\hbox{$\stackrel{{\displaystyle
\longrightarrow}}{_{x^2\rightarrow \infty}}$}}\ {Z_L(\m)\over x^2}
f^{acd} f^{bce} \gb^d \gb^e 
\eqn{Goldstonegp}
and of the $\OO^a$-propagator 
\eq
\vev{\OO^a(x)\ \OO^b(0)}\ {\raise-.7ex\hbox{$\stackrel{{\displaystyle   
\longrightarrow}}{_{x^2\rightarrow \infty}}$}}\ -{Z_L(\m)\over x^2}
f^{acd} f^{bce} \gb^d \gb^e
\eqn{OOcorel}
 also is a natural consequence of the massless Goldstone excitations. In the
case of covariantly quantized  gauge theories this coupling to
Goldstone excitations is unfortunately not as
inevitable as in a theory with a positive definite Hilbert space.

The residues in\equ{Goldstonegp} and\equ{OOcorel} are related by 
the  $\ts$-symmetry\equ{newbrs}, which implies that 
\eq
\vev{\ts\cb^a(x) \ts\cb^b(y)} =-\vev{\cb^a(x)\ \ts\ts\cb^b(y)}
=\a^{-1} \d^{ab} \d^4(x-y) 
\eqn{detalpha}
In gauges $\d=0$ this is the Slavnov-Taylor identity which determines the
longitudinal part of the gluon propagator completely. In general
gauges with $\d\neq 0$, \equ{detalpha} only relates connected Green's
functions and in particular requires the absence of Goldstone
excitations in the combination
\eqa
0=\a^2\vev{\ts\cb^a(x) \ts\cb^b(0)}_{x^2>0}&=&\left[\vev{(\OO^a(x)-\vev{\OO^a})\
(\OO^b(0)-\langle\OO^b\rangle)} +\vev{\pa\ipr A^a(x)\ \pa\ipr
A^b(0)}\right.\cr
&&\kern-7em +\left.\vev{(\OO^a(x)-\vev{\OO^a})\
\pa\ipr A^b(0)} +\vev{\pa\ipr A^a(x)\
(\OO^b(0)-\vev{\OO^b})}\right]_{x^2>0}
\eqan{absence}
for $x^2>0$.

The color structure of\equ{GoldstoneDA} leads to a cancelation of the
long range parts of the last two terms in\equ{absence}. 
The long range correlations of the first two terms  also
have to cancel for\equ{absence} to hold. Note that for $\d\neq
0$, \equ{absence} does not demand vanishing residue $Z_L$, 
because the metric of the Hilbert space of the covariant gauge theory is not
positive definite. 

States of negative norm in 
covariant gauges however also
prevent us from proving in the usual way
that $Z_L$ in \equ{Goldstonegp} does {\it not}
vanish. A non-trivial  cancelation of the Goldstone
poles between the first two terms of\equ{absence} however only
requires that the vertex 
$\Gamma^{[ab]}_{\OO\,\pa\ipr A}$
between the scalar $c \cb$-mode and longitudinal gluons 
with an antisymmetric color structure does not vanish for
$p^2\rightarrow 0$. We perturbatively verify that this is the case 
in the broken phase of the $SU(2)$-model in Appendix~C. 
The one-loop calculation of this vertex  gives  
\eq
\Gamma^{[ab]}_{\OO\,\pa\ipr A}(p^2\rightarrow 0) = \e^{abc}\gb^c\frac{3
g^2\d^2}{32\pi \k^2}  
\eqn{vertex}
Note that this dimensionless antisymmetric coupling arises only when
the global $SU(2)$ symmetry is spontaneously 
broken. A vanishing of the full $\OO\,\pa\ipr A$-vertex at zero
momentum and weak coupling would require a cancelation  of this
leading contribution from
non-perturbative corrections that are of order $1/g^2\propto 
\ln(\Lambda^2/\m^2)$. Such a dependence on the
renormalization scale $\m$
is generally induced only by divergences. Since the
vertex\equ{vertex} is not present in the 
action\equ{effaction} and the model is renormalizable, the
$\OO\,\pa\ipr A$-vertex antisymmetric in color is finite to all
orders.  We thus do not expect a cancelation of the lowest order
vertex\equ{vertex} at weak coupling.

The color structure of\equ{vertex} guarantees
the cancelation of Goldstone poles
in the Ward identity\equ{absence}. Since the
adjoint representation is the only one common to the antisymmetric
tensor product of two adjoint representations and the symmetric tensor
product of any number of adjoint representations, the corresponding
vertex for $SU(n)$ can only be proportional to 
$f^{abc}P^c(\gb)$ , where $P^c(\gb)=p_1\gb^c +p_2
d^{def}\gb^d\gb^e d^{cfg}\gb^g +\dots$ is an odd and totally symmetric
polynomial in $\gb$ in the adjoint representation of the group. That
the color structure of the Goldstone 
contribution to  $\vev{\pa\ipr A(x)\
\pa\ipr A(0)}$ is of the form\equ{Goldstonegp} then follows
from\equ{GoldstoneDA} and 
Bose-symmetry. The real, symmetric and positive semidefinite 
$(n^2-1)\times (n^2-1)$ matrix 
$M^{ab}=f^{adc}\gb^c f^{bde}\gb^e$ has exactly $(n-1)$
vanishing eigenvalues when $\gb$ breaks the $SU(n)$-group to
$U(1)^{n-1}$. Thus the $n(n-1)$ positive eigenvalues of $M^{ab}$
and the number of expected Goldstone modes correspond. Of some importance
for the next 
section  is  that the vertex $\Gamma_{\OO\,\pa\ipr A}$ is of order
$n^0$ for large $n$, since the
expectation value $\vev{\gb^a}$ is of order $n^0$ and insertions
of $\vev{\gb}$ therefore do not change the leading order in $n$ of a
diagram. The one-loop diagram shown in Fig.~5, which we evaluated for
$SU(2)$ in Appendix~C,  is
planar and thus  of order $n^0$ in  $SU(n)$. 

Finally note that the critical exponent of the residue $Z_L(\m)$
in\equ{Goldstonegp} is given by the anomalous dimension of the gluon
field of order $g^2$, since the anomalous dimension of $\gb$ is of
order $g^4$ in CCG.

Although indicative, we are well aware that these arguments for a Goldstone  
contribution to the longitudinal gluon propagator
in CCG are not conclusive. In the next section we
therefore  present further evidence for  a long range behavior of the {\it
longitudinal} gluon propagator of the form\equ{Goldstonegp}: assuming
that the dominant  
long-range contribution to a Wilson loop in CCG
comes from this Goldstone contribution to the longitudinal gluon
interaction, the expectation
value of  large Wilson loops is found to decay exponentially with the enclosed
area in the planar gluonic theory. 
  
\section{The Wilson loop in CCG}
The order parameter of confinement proposed by Wilson\cite{wi74} is the
expectation value of the non-local functional $W(\CC)$   
\eq
W(\CC):=\tr \PP \exp\left[g \oint_{\CC} dx_\m A_\m(x)\right]
\eqn{wilsondef}
where $\PP$ denotes path ordering and $\CC$ is a generic smooth
path. In a confining gluonic theory 
($n_f=0$) the expectation value of\equ{wilsondef} should decay exponentially
with the area of the minimal surface enclosed by $\CC=\pa D$
\eq
\ln\vev{W(\pa D)}\ {\raise
-.7ex\hbox{$\stackrel{\displaystyle\longrightarrow}{{\scriptstyle{\rm 
Area}(D)\rightarrow \infty}}$}}\  -\s{\rm Area}(D)
\eqn{vevW}
where the constant of proportionality $\s$ is the string tension. This
should in particular be 
the case for discs $D$ with large Euclidean radius $R$. 

$\vev{W(\pa D)}$ is a gauge- as well as RG-invariant observable for any
$SU(n)$ gauge theory. It suffices to show that the string
tension $\s$ does not vanish for large $n$, i.e. for the subset of planar
diagrams\cite{tH74} in the expansion of the Wilson loop. A 
cancelation  between planar and non-planar contributions to the
expectation value precisely for $n=3$, the gauge group of interest,
would be unfortunate. We will therefore only study planar
contributions to the expectation value\equ{vevW} of circular Wilson
loops such as the one shown in Fig.~2.
\vskip .5cm
\hskip 4cm\psfig{figure=sceleton.ps,height=1.3in}
\nobreak\newline
{\small\baselineskip 5pt 
\noindent Fig.~2: A planar gluonic contribution to a circular
Wilson loop.}    

Although the leading order in $n$ of $\vev{W(\pa D)}$ is gauge invariant,
the contributions from planar connected gluonic Green's functions
to a diagram such as Fig.~2 {\it are} gauge dependent {\it
  individually}. It is therefore  possible, that some of these
contributions to the expectation value are 
negligible compared to others in  {\it certain} gauges when the area
of the loop is large.  In the
preceding sections we argued that the longitudinal gluon
propagator in CCG is of particularly long range due to Goldstone
modes\footnote{With $n_f=0$ the two CCG are at gauge parameters
  $\d^\pm=\frac{11\pm\sqrt{22}}{9}$ and  
$\a^\pm=\frac{13(13\pm2\sqrt{11})}{81}$}. The planar connected Green's
functions contributing to large Wilson loops naturally
fall into two categories in CCG: i) those which are
dominated by the Goldstone poles  at long range and ii) those
which aren't. Contributions of type i) at large distances can be
regarded as longitudinal gluon exchanges with an effective
coupling $G(\m)$ to the loop. $G(\m)$ coincides with the
gauge coupling $g(\m)$ in lowest order of the  weak coupling expansion
only and is also of order $1/\sqrt{n}$. (The order $g^3(\m)$
corrections to $G(\m)$ are obtained in Appendix~D. One finds that $G(\m)$,
the effective coupling of a longitudinal gluon to the Wilson
loop, and the perturbative coupling $g(\m)$ differ in their dependence
on the renormalization scale $\m$.) In  critical gauges
contributions of type i)  give rise to a long range
interaction 
\eq
v_\mn^{ab}(x^2\sim\infty) = \left.G^2(\m) 
\vev{A_\m^a(x)\  A_\n^b(0)}\right|_{Goldstone}
\eqn{interac}
from the exchange of $n(n-1)$ Goldstone bosons. Because the $SU(n)$
group is broken to $U(1)^{n-1}$, the number of Goldstone bosons is of
order $n^2$. To show that the interaction\equ{interac} is also of
leading order in $n$, however requires that their average residue 
does not vanish in the limit of large $n$.   As remarked earlier, it
is evident from the action $\equ{effaction}$ that insertions
of $\gb$ are of order $n^0$ and thus 
do not change the leading order in $n$ of a diagram. Thus the
effective potential\equ{veff} is of order $n$, but its minimum,
$\vev{\gb}$ as well as the scale $\k$ are of 
order $n^0$. Since all eigenvalues  of $\vev{\gb}$ 
differ and each is of order $n^0$, the sum of their squares therefore
grows like $n$, that is   
\eq
\gb^a\gb^a=-2\tr\gb^2=2\sum_{i=1}^n\lambda_i^2=O(n)
\eqn{powern}
because at most one of the eigenvalues vanishes.
To leading order in $n$ we can therefore
replace the color dependence of\equ{Goldstonegp} by the average
\eq
\vev{\pa\ipr A^a(x)\ \pa\ipr
A^b(0)}{\raise-.7ex\hbox{$\stackrel{{\displaystyle
\longrightarrow}}{_{x^2\rightarrow \infty}}$}}\ {Z_L(\m)\over
x^2}\d^{ab}\frac{\gb^c\gb^c}{n} +O(1/n)
\eqn{Goldstonegpn}
Due to \equ{powern}, the Goldstone contribution to the longitudinal
propagator is of
$O(1)$, if the residue $Z_L$ of the Goldstone modes is of order
$n^0$. The previous argument that the coupling $Z_L$ does not vanish
 is also valid in leading order of $n$ (since the Ward
 identity\equ{Goldstones}
is of leading order and the leading contributions to the vertex
$\Gamma_{\OO\, \pa\ipr A}^{[ab]}$ shown in Fig.~5 are planar).  As 
far as color counting is concerned, the exchange of Goldstone
modes in CCG therefore contributes to the Wilson loop in leading order
of the $1/n$ expansion.

Integrating\equ{Goldstonegpn} we obtain the long-range behavior of the
longitudinal gluon propagator to $O(1)$
\eq
\vev{ A_\m^a(x)\ A_\n^b(0)}{\raise-.7ex\hbox{$\stackrel{{\displaystyle
\longrightarrow}}{_{x^2\rightarrow \infty}}$}}\
-Z_L(\m)\d^{ab}\frac{\gb^c\gb^c}{32 n} \pa_\m\pa_\n x^2\ln(b(\m) x^2)
\eqn{gplr}
where we have assumed the least singular behavior for $x^2\rightarrow
\infty$ compatible with\equ{Goldstonegpn}. The
integration constant 
$b(\m)$ cannot be determined at this stage. We will numerically find
that $b(\m)$ is  
related to $Z_LG^2$ if the physical spectrum is 
unitary. At present we only remark that changing the
parameter $b$ alters the propagator by a constant and therefore
changes its long range  behavior. The Goldstone pole in the
correlator\equ{Goldstonegpn}, which was inferred 
from the Ward identities, only determines the {\it leading} long-range
behavior of the propagator which does not depend on the associated scale
$b$. Note however that an explicit solution of the model, would
give the
long range part of the gluon propagator up to a physical scale and
determine the associated dimensionless parameter $Z_L/b$ uniquely in
terms of the dimensionless coupling constants.

Inserting\equ{gplr} in\equ{interac} we arrive at a parameterization of
the long range 2-point interaction in leading order of $n$ 
\eq
v_\mn^{ab}(x^2\sim\infty) = \frac{\d^{ab}}{n} K^2 \pa_\m\pa_\n
x^2\ln(s K^2 x^2/e^2) 
\eqn{interac2}
in terms of $K$ and $s$ which are related to $G^2 Z_L$ respectively
$b/(G^2 Z_L)$.
In\equ{interac2} $s$ is a dimensionless parameter  and $K$ is an inverse
length. In Appendix~D we verify that
these parameters approach renormalization group invariants for
$\m\rightarrow\infty$. To establish this,  we consider the {\it
perturbative} planar 1-loop corrections of Fig.~6 to the effective long-range 
exchange\equ{interac2} within a Wilson loop. The explicitly
calculation of Appendix~D shows that the UV-divergences
associated with an evaluation of diagrams~6a+b 
precisely cancel the renormalization factor for the propagator and
coupling to order $g^2$. No UV-subtractions of order $g^2$ are
therefore necessary to render the contribution to the expectation
value of the Wilson loop from effective planar longitudinal 2-point
exchanges finite. The interaction\equ{interac2}
therefore does not depend on the renormalization scale $\m$ as
$\m\rightarrow \infty$. 

In Appendix~E we prove that planar contributions to the Wilson loop  
generated by an effective 2-point longitudinal interaction $v$  are resummed
by a nonlinear integral equation. In the case of a circular Wilson
loop of radius $R$ this integral equation is of the relatively simple
form 
\eq
W_R(\t)=\one+\int_0^\t d\t^\prime \int_0^{\t^\prime} d\t^{\prime\prime} 
v_R(\t^{\prime\prime}) W_R(t^{\prime\prime}) W_R(\t^\prime-\t^{\prime\prime})
\eqn{eqWR}
In\equ{eqWR} the color traces have been performed and the expectation
value of the circular Wilson loop $W(\pa D_R)$  of large radius $R$
with the above approximations is
\eq
\vev{W(\pa D_R)}\ \stackrel{R\sim\infty\atop\sim}\ n W_R(2\pi)\ .
\eqn{relWDtoWR}
The effective interaction $v_R(\t)$ in\equ{eqWR} is related
to\equ{interac2} by
\eqa
v_R(\t) &=&\left. \pad{x_\m}{\t_x} \pad{y_\n}{\t_y} \frac{1}{n}\tr t^a t^b
v_\mn^{ab}(x(\t_x)-y(\t_y))\right|_{\t=\t_x-\t_y}+O(1/n^2)\cr
&=&\half \pad{^2}{\t^2} (KR)^2 \sin^2(\t/2)\ln(s (KR/e)^2 \sin^2(\t/2))\cr
&=& (KR)^2 \left[1+ \cos(\t)\ln(s (KR)^2 \sin^2(\t/2))\right]
\eqan{vRrel}
since the distance $(x-y)^2=4 R^2\sin^2((\t_x-\t_y)/2)$ of two points
on the loop depends only on the radius $R$ and the angular difference
$\t=\t_x-\t_y$ between them. The interaction\equ{vRrel} depends
parametrically on the dimensionless area $(KR)^2$ of the loop and the 
dimensionless parameter $s$.

We numerically solved the nonlinear integro-differential equation
\eq
\pad{}{\t}W_R(\t)=\int_0^\t d\t^\prime\, v_R(\t^\prime) W_R(\t^\prime)
W_R(\t-\t^\prime)\,,\quad W_R(0)=1
\eqn{diffintegro}
which is equivalent to\equ{eqWR}.
For small values of $\t\ll (KR)^2$ \equ{diffintegro} was solved
iteratively and we then used these starting values to extend the
solution to $\t=2\pi$ by a predictor-corrector algorithm based on the
Adams-Bashforth-Moulton scheme.  The numerical integrations
where performed with a
modified Rhomberg algorithm\cite{nr86}. We found the predictor-corrector
method appropriate in this case, since the integral in\equ{diffintegro}
depends on all previous values for the function $W_R$. $W_R(\t)$
itself is a reasonably
smooth function  which oscillates a few times in the interval
$[0,2\pi]$. 
Special care was taken to correctly take into account the logarithmic
(but integrable) singularities of $v_R$ at $\t=0$ and $\t=2\pi$. The 
algorithm is exceptionally stable and we quote results
which we believe\footnote{We used up to $10^4$ integration points in
the interval $[0,2\pi]$ and the predictor-corrector algorithm is of
fourth order.}  to be accurate to $\sim 10^{-12}$. The results of these
numerical simulations are summarized in Fig.~3. For a particular value
 $s= s_{crit.}=1.195$, we numerically find that the
expectation value of the Wilson loop decreases exponentially with the  
dimensionless area $(KR)^2$ over $12$ orders of magnitude. The results
for the Wilson loop with neighboring values $s=1.17$ 
and $s=1.25$ are also shown in Fig.~3 for comparison. We observe that for
$s=1.17<s_{crit.}$ 
the logarithm of the Wilson loop  decreases faster than linear with the
area. Such a behavior of the expectation value is excluded by
causality\cite{se80}. For $s=1.25>s_{crit.}$ we also observe a violation
of causality -- for large values of the area, the expectation value
increases again. This behavior is more marked at even larger
values of $s$, as can be seen for $s=2.0$ in Fig.~3. At $s=s_{crit.}$ the
oscillations which eventually cause the expectation value of the
Wilson loop to become negative are of order $10^{-13}$ and thus well
below the numerical accuracy of the algorithm. 

The results summarized in Fig.~3  confirm our expectation that the
long range interaction\equ{interac2} is only consistent with
unitarity at a particular value of the dimensionless parameter $s$. Our best
numerical determination of this critical value for the parameter is
$s=s_{crit.}\sim 1.195\pm .01$.  A
full solution of the model would {\it predict} a value for the
dimensionless parameter $s$ in the asymptotic interaction and
furthermore relate
the scale $K$ to $\L_{ASP}$. With the assumption that the
approximations made in the evaluation of the expectation value of
large Wilson loops are reasonable in CCG, the numerical evidence
suggests that the string tension does not vanish and gives an estimate
for $s_{crit.}$  if we require unitarity. Note that 
$s_{crit.}$ is a dimensionless parameter which would be a prediction
of the gauge theory,
whereas $K$ is a scale, which  essentially has to be  determined
experimentally. If the 
residue $Z_L$ in\equ{Goldstonegp} and the expectation value
$\gb$ can be related to $\Lambda_{ASP}$ in a two-loop calculation, 
a quantitative  estimate of $\L_{ASP}/K$ eventually could be possible
within the framework of perturbation theory. 

\epsfxsize=7.0 in
\hspace{-0.5in}\epsfbox{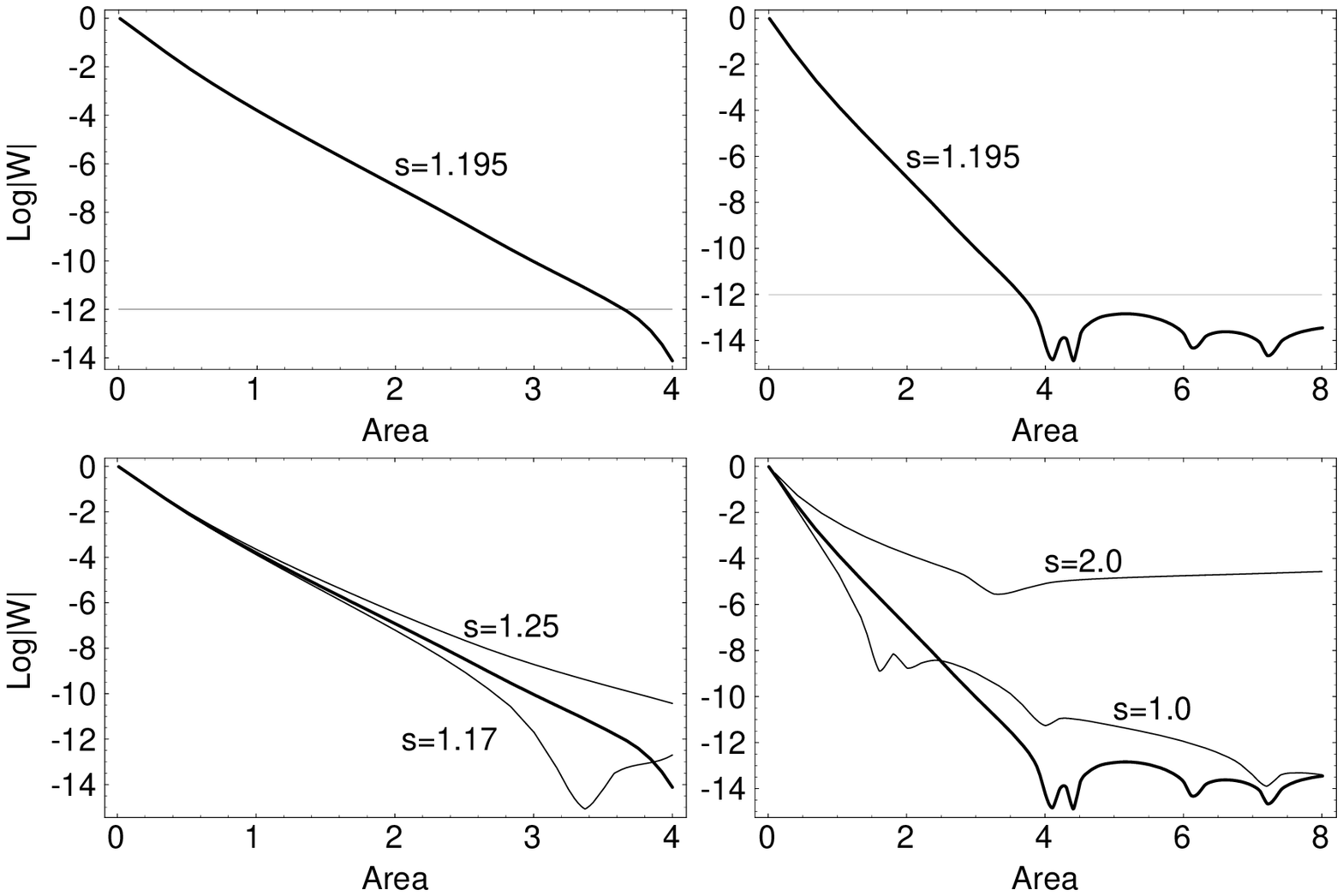}
\nobreak\newline
{\small\baselineskip=5pt
\noindent Fig.~3: The expectation value of large circular Wilson loops as a
function of the enclosed dimensionless area $(KR)^2$ in the long range
approximation of CCG. \nobreak\newline
{\it Top-left}: The logarithm of the expectation
value for $s=1.195$ in the range $(KR)^2<4$; the expectation
value is positive in this range. The line at $\log(\vev{W})=-12$ is an
estimate of the accuracy of the numerical solution.\nobreak\newline
{\it Top-right}: same as {\it top-left} but for the greater range
$(KR)^2<8$;  the
plot shows the logarithm of the absolute value of the 
expectation value, which changes sign at the sharp dentures for
$(KR)^2 >4$. We attribute these oscillations of order $10^{-13}$ in
the expectation value to numerical noise of the algorithm.\nobreak\newline
{\it Bottom-left}: The logarithm
of the absolute value of the expectation value of the Wilson loop in
the range $(KR)^2<4$ for parameter $s=1.195$ (bold curve) as
well as $s=1.17$ and $s=1.25$; for $s=1.17$ the  
logarithm of the expectation value decreases faster than linearly with
the area and  violates unitarity. \nobreak\newline
{\it Bottom-right}: Same as {\it
bottom-left} for a wider range of the parameter $s$ and for
$(KR)^2<8$. The result for $s=2.0$ shows a violation of
unitarity at the level of $10^{-4}$ which could be interpreted as due
to a low-mass state of negative norm.}

Our numerical studies also indicate that the Goldstone pole
in\equ{Goldstonegp} is essential for 
an exponential decrease of the Wilson loop in this approximation. The
logarithmic dependence on the distance of the effective
interaction\equ{interac2} and\equ{vRrel} is a direct consequence of,
and implies, a Goldstone pole 
in\equ{Goldstonegp}. We numerically investigated the
effect of replacing the logarithm
in\equ{vRrel} by a constant and found that this immediately leads to a
non-unitary dependence of the expectation value of the Wilson
loop. Our numerical results therefore depend sensitively on the
asymptotic effective 
interaction\equ{vRrel} and are not a generic consequence of the
non-linear integral equation\equ{eqWR}. Partly due to this
sensitivity however, we presently do not analytically understand 
the solution to this nonlinear integro-differential equation. An
analytic, or approximate, solution to\equ{diffintegro} for large
values of $(KR)^2$ would provide a better
understanding of this confinement mechanism in CCG.

\section{Discussion} 
Since quarks have not been observed as asymptotic states, confinement
is one of the central issues in $SU(n)$ gauge theories without
fundamental scalars. On 
the other hand, these models are asymptotically free and perturbation
theory in conjunction with the RG-equation is a powerful tool in
their investigation. It was only recently observed that a translationally
invariant  BRST-quantization
of these theories on compact space-time manifolds 
introduces a non-trivial moduli space\cite{bs97,brs96} because the
BRST-symmetry has to be realized equivariantly with respect to global
gauge transformations. In this work we
found non-trivial fixed points on this moduli space in CCG,
which respect the equivariant BRST-symmetry but break  the global
$SU(n)$-symmetry spontaneously to $U(1)^{n-1}$. The associated
Goldstone bosons are 
$c\cb$-bound states in the adjoint representation, which we argued
lead to an IR-singular long distance behavior of the {\it
  longitudinal}  gluon correlation of the form\equ{Goldstonegp}. Our
numerical results for the Wilson loop 
indicate that this long range behavior of the longitudinal gluon
propagator confines.

This scenario is however only realized in particular covariant
gauges and only for $n_f\leq n$ quark flavors. In gauges other than CCG, the
behavior  of the Wilson loop is presumably similar,  but the
dynamics leading to confinement 
cannot be described in this way. More specifically, the relevant unphysical
degrees of freedom that give a confining behavior of large Wilson loops in
other covariant gauges are field configurations which cannot
be described by the limited moduli-space at our disposal. The
situation is somewhat analogous to atoms in QED:
the spectrum of an 
atom is gauge invariant, but its qualitative description in gauges other than
Coulomb gauge is a  non-trivial matter. In Coulomb gauge, the
unphysical degrees of freedom  conspire to give a potential which 
describes the spectrum qualitatively  and  radiative corrections
are computable and small. We believe that  gauge freedom can
also be exploited advantageously in $SU(n)$, but with the important
difference that these non-abelian  models are asymptotically
free. Perturbation theory is valid only at high energies and we rely 
on spontaneous symmetry breaking to relate perturbative information to
low energy dynamics.  Within our limited moduli space, we can verify
such a symmetry breakdown only in 
CCG and only for $n_f\leq n$. As far as we know, the expectation value of
some composite scalar not 
described by our moduli space, would show that $SU(n)$ is broken to
$U(1)^{n-1}$ also in covariant gauges other than CCG, and the corresponding
Goldstone excitations could similarly lead to confinement. The limitation
of our perturbative approach is that we cannot access the 
relevant composite scalars in gauges other than CCG. In non-covariant gauges
the confinement mechanism is probably quite different and we have no
suggestion how to perturbatively gain information about the low energy
dynamics of the theory in this case.

The outstanding (and new) feature of the proposed confinement
mechanism in CCG is the Goldstone pole in $\vev{\pa\ipr
  A(x)\ \pa\ipr A(0)}$, which implies that the {\it longitudinal}
gluon interaction is proportional to $\sim 1/k^4$ for small
momentum transfers. In conventional covariant gauges, the BRST-symmetry
constrains the longitudinal propagator to be proportional to $1/k^2$;
a $1/k^4$ behavior of the longitudinal propagator however does not
violate Ward-identities in the 
extended covariant gauges with quartic ghost interaction ($\d\neq
0$). We argued that  perturbation theory is consistent with a dynamical
spontaneous breaking of the global $SU(n)$-symmetry for certain
critical gauges due to $ghost-antighost$-condensation in the
adjoint. Remarkably, this spontaneous symmetry breaking of the gauge
fixed theory does not give rise to a mass for the longitudinal degrees of
freedom of the gauge field via the Higgs mechanism  and the
vertex\equ{vertex} we obtain for the coupling between 
these modes does not vanish for $p^2\rightarrow 0$.  This is probably
because the coupling of the longitudinal gauge field to the massless
(composite) $c\cb$-Goldstone modes is not constrained by gauge
symmetry. 
 
We presented numerical 
evidence that the observed behavior of the {\it longitudinal} propagator
in non-abelian theories would confine in the sense of Wilson\cite{wi74}.
An appealing feature of this scenario is that the IR-singular behavior
of the {\it longitudinal} exchange does not lead to a strong ``Van
der Waals'' force between two distinct Wilson loops at large
distances, because the interaction  is a total derivative in both loop
parameters.  
For sufficiently small loops, for which all exchanges within the loops
can be neglected, the force between two Wilson loops due to longitudinal
exchanges is therefore proportional to the physical size\footnote{Since the 
  contribution from the exchange of one and two longitudinal gluons 
  vanishes, the force is at least proportional to the
  sixth power of the physical size of either Wilson loop}
of either loop.  The Van der Waals force from longitudinal exchange
between {\it local} physical observables, for which this approximate
evaluation of the long range interaction is valid  in CCG, is thus
seen to vanish.

Let us finally point out that the symmetry breaking we have
discussed gives {\it quantitative} corrections to conventional perturbative
results. These can in principle be calculated and experimentally
verified. Of particular interest are perhaps power
corrections to physical correlation functions at large Euclidean momentum
transfers. These power corrections are described by vacuum
expectation values of composite operators in
the operator product expansion of Wilson\cite{wi69}. At present these
expectation values are determined experimentally or are obtained
from lattice simulations. In CCG such power corrections  arise
naturally  in the loop expansion due to the  expectation values of the moduli.

\begin{center}{\bf Acknowledgments}\end{center}
We have greatly profited from extensive discussions with D.~Zwanziger
and L.~Baulieu. This work could not have been finished without their
support. We would like to thank D.Kabat and M.Porrati for explaining
some aspects of confinement in supersymmetric gauge theories.
A.R. enjoyed the hospitality  
of the LPTHE where this work was completed and was partially
supported by  a Margaret and Herman Sokol Research Fellowship. 

\begin{appendix}
\section{The partition function $\vev{\one}$}
We here adapt the proof\cite{bs97} that the
partition function of the TQFT of the gauge group  does not vanish  
to our case. Consider the partition function
\eq
\ZZ[A]=\vev{\one}_A =\int [dU] [dc] [d\cb] [db] d\f d\s d\sb
d\gb  d\g\  e^{S_{A}} 
\eqn{norm}
of the equivariant TQFT  on the gauge group with action  
\eq
S_A=s W_{GF}[A^U]
\eqn{actionA}
where $W_{GF}[A^U]$ is the gauge fixing functional \equ{wgf} with $A$
replaced by
\eq
A_\m\rightarrow A^U_\m= U^\dagger A_\m U + \frac{1}{g} U^\dagger \pa_\m U
\eqn{replace} 
$A_\m$ in\equ{norm} is a
background connection  that identifies a particular
orbit. $U(x)\in SU(n)$ 
is a local gauge transformation and the action of the BRST-operator
$s$ on the fields is 
defined as in\equ{sdef} with the replacement\equ{replace}. The action
of $s$ on the background connection $A$ and the gauge transformation
$U$ individually is given by
\eqa
s A_\m(x) &=& 0\cr
s U(x) &=& U(x)(g c(x)+\o)
\eqan{stop}
$s$ thus effects an infinitesimal variation of the gauge group element
$U$. It is straightforward to show that $s$ is nilpotent ($s^2=0$). 
Because the action\equ{actionA} is BRST-exact, \equ{norm} is the
partition function of an equivariant TQFT of Witten type\cite{bi91} on the gauge group. Standard arguments of
TQFT show that \equ{norm} does not depend on the parameters $\a$ and
$\d$ in\equ{wgf} and that \equ{norm} is proportional to the generalized Euler
character of the space 
\eq
\EE[A]:=\{U: U(x)\in SU(n), \pa\ipr A^U=0\}/SU(n)
\eqn{space}   
of solutions to the Landau gauge condition $\pa\ipr A=0$ modulo global
(right-hand) gauge transformations. 

On the other hand the partition function $\vev{\one}$ of the
equivariantly gauge fixed $SU(n)$ theory  is proportional to 
\eq
\vev{\one}=\int [\DD\psi\DD\bar\psi\DD A] \ZZ[A] e^{S_C}
\eqn{partition}
where $S_C$ is the classical gauge invariant
action\equ{classaction}. Since the gauge symmetry is not anomalous, the
change of variables 
\eq
A^U\rightarrow A\ ,\qquad U\psi\rightarrow \psi \ ,\qquad \bar\psi
U^\dagger \rightarrow \bar\psi
\eqn{change}
decouples the $U$-integration which is assumed to be normalized.
The remaining path integral is just the one we are considering and the
BRST-symmetry of the TQFT becomes the BRST-symmetry\equ{sdef} of the
model.  

We show that $\ZZ[A]$ is a non-vanishing constant that does not
depend on the background connection $A$. Since we are defining the
model on a finite torus, 
an explicit mode expansion is possible. The Landau gauge condition is
associated with fixed points of the Morse potential\cite{zw94}
$\VV_A[U]\in \RR_+<\infty$ 
\eq
\VV_A[U] = \int_\MM \tr A^U\ipr A^U
\eqn{Morse} 
which is invariant with respect to global right-hand $SU(n)$
transformations. The generalized Euler characteristic is the (signed)
sum of the Euler characters $\chi$ of the finite dimensional fixed point
manifolds of\equ{Morse} modulo $SU(n)$
\eq
\chi(\EE[A])=\sum_{\left.\d \VV_A[U]\right|_{U\in \MM_i}=0}\pm\chi(\MM_i/SU(n)) 
\eqn{gEc}
where the sign depends on the number of negative eigenvalues of the
Hessian at a point of the fixed point manifold\cite{bi91}. It is clear
that this 
generalized Euler characteristic is constant under continuous
deformations of the Morse potential $\VV_A$. Since the space of 
$SU(n)$ connections on a (finite) torus is connected,
$\ZZ[A]$ is thus a constant independent of $A$ and it suffices to show that
$\ZZ[A=0]$ does not vanish.
To simplify matters, we only show that the generalized Euler
characteristic of $\EE[A=0]$ does not vanish for an $SU(2)$ gauge
group on a torus.
 
The Morse potential $\VV_{A=0}$ for $A=0$ is also
symmetric with respect to {\it left} multiplication by a (constant) group
element: $U(x)\rightarrow g_L U(x)$. Consequently, if $U(x)$ is a
critical point of the potential\equ{Morse} at $A=0$, 
\eq
\pa\ipr(U^\dagger \pa U)=0
\eqn{classical0}
then so is
\eq
U(x,g_L,g_R) = g_L U(x) g_R,  \qquad \forall g_L,\,g_R\in SU(n)
\eqn{mods}
Some of these solutions however belong to the same equivalence class
modulo right multiplication by global $SU_R(n)$ transformations. We obtain the
moduli space of these (right equivalence classes of) solutions
to\equ{classical0} by noting that left
multiplication of $U(x)$ by $g_L\in SU(n)$ is the same as right
multiplication by  
\eq
g_R(x)=U^\dagger(x) g_L U(x)
\eqn{eqgr}
$g_L U(x)$ therefore belongs to the equivalence class modulo right
multiplication of $U(x)$ by constant group elements only if $d g_R(x)=0$, i.e.
\eq
[U(x)d U^\dagger(x),\, g_L]=0
\eqn{equivlr}
Thus left multiplication of $U(x)$ by any $g_L\in SU(n)$ belonging
to the subgroup which commutes with $U(x) d U^\dagger(x)$ is
redundant. For an $SU(2)$ gauge group there are  only three possible
subgroups to consider: 
\begin{itemize}
\item[1)] $g_L\in  SU_L(2)$ satisfy\equ{equivlr} $\Rightarrow
 \chi(SU_L(2)/SU_L(2))=1$ 
\item[2)] $g_L\in U(1)\subset SU_L(2)$ satisfy\equ{equivlr} $\Rightarrow
\chi(SU_L(2)/U(1)\simeq S_2)=2$
\item[3)] $g_L\in \{\one, -\one\} \subset
SU_L(2)$ satisfy\equ{equivlr} $\Rightarrow
\chi(SU_L(2)/\{\one,-\one\}\simeq SO(3))=0$
\eq
\eqn{cases}    
\end{itemize}
Note that case 1) implies that $U(x) d U^\dagger(x)=0$ and therefore
corresponds to the equivalence class of the identity (modulo right
multiplication). There is only  {\it one} such class. For an $SU(2)$
gauge group, case 2) only occurs if $U(x)d U^\dagger(x)=
d\t(x)$ is an {\it abelian} connection. Since $U(x)$  should
furthermore satisfy\equ{classical0}, we can conclude that this
connection maps the 1-cycles of the torus to $U(1)$. A typical
transformation of this kind on the torus  has the form
\eq
U(x)=e^{a_\m x_\m}\,, {\rm \ with}\ [a_\m,\,a_\n]=0 
\eqn{specialU}
with a constant abelian connection $a_\m$, whose components on a
symmetric torus of extension $L$ are 
\eq
a_\m=2\pi i\hat a\ipr\vec{\s} k_\m/L,\ k_\m\in\integers
\eqn{conna} 
where the integers $k_\m$ are the winding numbers of the mapping of the
torus to the $U(1)$ subgroup  and $\hat a$ is a unit vector which is the
same for all components. Note that the corresponding constant abelian
connections 
\eq
A_\m(x)=\frac{1}{g} U^\dagger \pa_\m U=\frac{1}{g} a_\m
\eqn{constcon}
are the Gribov copies of $A=0$ on a torus we discussed in
section~2.1. They come in
pairs $(a,-a)$ that are related by a discrete $\integers_2$ subgroup of the
$SU_R(2)$-transformations. A direct consequence is that the
expectation value $\vev{a_\m}$ vanishes for the equivariantly
gauge fixed $SU(2)$ theory on a 
torus. 

{F}rom\equ{cases} we obtain that the moduli space
of solutions to\equ{classical0} of an $SU(2)$ theory has the
topological structure
\eq
\EE_{A=0}=SU_R(2)\times\left[\one + S_2\times\FF+ SO(3)\times
\widetilde\FF\right]  
\eqn{structure}
In other words $\EE_{A=0}/SU(2)$ can essentially be described as a
single point and a collection of two- and three-dimensional
spheres\cite{gr78, vb95}. 
Although we do not know the Euler numbers associated with the
topological spaces  $\FF$ and $\widetilde\FF$, the
structure\equ{structure}, together with $\chi(S_2)=2$ and
$\chi(SO(3))=0$ suffices to obtain that on a torus  
\eq
\chi(\EE[A=0]/SU(2))= odd\neq 0
\eqn{espace}
In the vicinity of $A=0$ the degeneracy with respect to left group
multiplication is in general lifted, but the signed sum of Morse
indices over the (then isolated) fixed points still gives the Euler
number\equ{espace}. Since the space of $SU(2)$ connections on a torus
is connected, we see that the partition 
function\equ{norm} of the TQFT is a constant independent of $A$ which
does not vanish.
To see that the partition function $\vev{\one}$ of the gauge fixed
theory\equ{partition} can then also be normalized, we observe that
the determinant of the Euclidean  Dirac-operator is positive for
non-vanishing quark masses in this vector-like theory.

\section{The one loop effective potential of $SU(n)$}
We here find the dependence of the 1-loop effective potential on the
moduli $\gb$ by extending the derivation of the effective
potential\cite{bs97} for an $SU(2)$ 
group to $SU(n)$. For reasons given in section~4, all moduli
except $\gb$ have vanishing expectation values. We take
space-time  to be  a symmetrical torus $L\times L\times \dots  
\times L$ of $D$ dimensions\footnote{Apart from the finite volume we
essentially  follow the procedure of ref.\cite{co73a} using
dimensional regularization and the MS-scheme}. The 
classical contribution to $V_{eff}(\gb)$ is the quadratic term
$-Tr\gb^2/\a g^2$ in the action\equ{effaction}. The $g$-independent term of  next order in the loop expansion is obtained by evaluating the infinite sum
of one ghost loop diagrams shown in Fig.~4. Crucial for the following
is the absence of a compensating gluonic one-loop contribution to the
effective potential, because the direct coupling $\int_\TT \frac{1}{\a
  g}\gb \pa\ipr A$ of $\gb$ to the longitudinal gluon field
in\equ{effaction} vanishes on a torus with periodic boundary
conditions for the gauge field.
\vskip .5cm
\hskip 4cm\psfig{figure=ghostloop1.ps,height=1.3in}
\nobreak\newline
{\small\baselineskip 5pt 
\noindent Fig.~4: Nonvanishing one-loop contribution to the effective
moduli action: a ghost loop with $2p
$ insertions of $\gb$-moduli.}  

Note that the antisymmetry of the structure constants $f^{abc}$
implies that only loops with an even number of 
$\gb$-insertions give a non-vanishing contribution to the effective potential.
The  loop with $2p$  insertions of the ghost $\gb$ 
is proportional to
\eq
C_p=\tr_{adj.} \hat\gb^{2p}
\eqn{colorfac}
where we have used the notation $\hat X$ 
to denote the
(anti-hermitian) matrix of the adjoint representation of $X$, $\hat
X_{ab}=f^{abc} X^c $.  

We want to express the trace\equ{colorfac} in terms of the eigenvalues of
$\gb$ in the fundamental representation. 
The trace of interest is given by 
\eq
\tr_{adj.}\hat{\gb}^{k}=-2\tr\,[[[\ldots [t^a,\underbrace{\gb]\ldots,\gb ],\gb ],\gb ]}_{k\,\,\mbox{times}}t^a
\eqn{tr}
where $t^a$ are the anti-hermitian generators of the fundamental
representation of $SU(n)$ normalized to $\tr t^a t^b=-\half\d^{ab}$. 
Using the Baker-Hausdorff formula the right hand side of \equ{tr} is written
\eq
\left.\frac{d^k}{d\a^k}\tr\,e^{-\gb\a}t^ae^{\gb\a}t^a\right|_{\a=0}
\eqn{tr1}
We choose the basis in which $\gb$ is diagonal:
$\gb_{kj}=i\lm_k\d_{kj}$ (no summation over $k$). The completeness
relation for the $t^a$'s, \mbox{$t^a_{ij}t^a_{lm}=-\half
\left(\d_{im}\d_{lj}-\frac{1}{n}\d_{ij}\d_{lm}\right)$}, gives:  
\[
\tr\,e^{-\gb\a}t^ae^{\gb\a}t^a=\frac{1}{2}-{\half}\tr\,e^{-\gb\a}\tr\,e^{\gb\a}
\]
The evaluation of the derivatives of this expression for $p>0$ leads
to the desired expression of $C_p$ in terms of the eigenvalues $\lm_i$ 
\eqa
C_p &=&-2\frac{d^{2p}}{ d\a^{2p}}\tr\,e^{-\gb\a}t^ae^{\gb\a}t^a|_{\a=0}=\sum_{k=0}^{2p}C_{2p}^{k}(-1)^k\tr\gb^k\tr\gb^{2p-k}\cr
&=&\sum_{k=0}^{2p}C_{2p}^{k}(-1)^{p+k}\sum_{i}^n\lm_{i}^k
\sum_{j}^n\lm_j^{2p-k}=2(-1)^p\sum_{1\leq i<j\leq n}(\lm_i-\lm_j)^{2p}
\eqan{eval1}

The contribution to the effective potential from a single loop with $2p$
insertions of the moduli is
\newcommand{\sumn}{\sum_{\{n_1\dots n_D\}\neq\{0\dots 0\}}}
\eq
F_D(p)= \frac{1}{2p L^D}\tr_{adj.}
\left(\frac{L^2\d\hat\gb}{(2\pi)^2}\right)^{2p} 
\sumn \left( n_1^2 +\dots+ n_D^2\right)^{-2p}  
\eqn{sumn}
The sum in\equ{sumn} extends over all the sets of integers $\{n_1\dots
n_D\}$ describing the complete set of modes with momenta $k_\mu=2\pi
n_\mu/L$ of the dynamical ghosts. The contribution from the constant
modes with $n_\mu=0, \ \mu=1,\dots,D$,  to\equ{sumn} is however
eliminated  by the integration over the ghosts $\sb$ and $\g$
in\equ{effaction}. 
$F_D(p)$ is therefore finite for 
$p>D/4$. The overall sign of\equ{sumn} is due to
ghost statistics. 

One can analytically continue to non-integer dimensions by casting $F_D(p)$
in integral form, 
\eq
L^D F_D(p)=  \tr_{adj.} \left(\frac{L^2\d\hat\gb}{(2\pi)^2}\right)^{2p}
\frac{1}{\G(2p+1)} \int_0^\infty dx\,x^{2p-1} 
\left[\left(f(x) \right)^D -1\right] 
\eqn{FD} 
where the function $f(x)$ for $x>0$ is the convergent sum,
\eq
f(x)= \sum_{n=-\infty}^\infty e^{-x n^2} 
\eqn{fdef}
Although there is no analytic expression for $f(x)$ at arbitrary
values of its argument, the reflection formula\cite{el89}
\eq
f(x)= \sqrt{ {\pi}\over{x}}\ f({  {\pi^2}\over {x}}) 
\eqn{reflection}
implies the asymptotic behavior,
\eqa
f(x\rightarrow\infty)&=& 1 + 2e^{-x} + O( e^{-4x})\cr
f(x\rightarrow 0) &=&\sqrt{\frac{\pi}{x}}(1 + O(e^{- {{\pi^2}\over x}}))
\eqan{asymptotic}

Since the eigenvalues of the commuting ghost $\gb$ can be treated as
real numbers 
\eq
v_{ij}^2=\frac {(\lm_i-\lm_j)^2\d^2}{(4\pi)^2\m^4}
\eqn{r}
is positive and the 1-loop contributions to the effective potential  can
be  summed over all values of $p\geq 1$. In\equ{r} we have introduced
the {\it finite} scale $\mu$ because we are 
interested in the $\mu L\rightarrow\infty$ limit and the eigenvalues
$\lm_i$ in this limit are measured in terms of the finite
renormalization scale $\mu$. 
 Summing over all one loop contributions and adding  the
tree-level potential $-\tr\gb^2/\a g^2=\sum_{i<j} (4\pi)^2 \m^4
v_{ij}^2/n\a \d^2 g^2 $  
gives
\eq 
V_{1-loop}=\sum_{1\leq i<j\leq n}\m^4 \frac{(4\pi)^2 v_{ij}^2 }{n\a \d^2 g^2}  
-4 L^{-D}\int_0^\infty \frac{dx}{x} \sin^2( v_{ij}x/2))
\left[\left(f(\pi x/(L\m)^2) \right)^D -1\right]   
\eqn{sumE} 
The asymptotic behavior of $f(x)$ for  $x\rightarrow 0$
determines the leading behavior of the integral in\equ{sumE} when
$L\m\rightarrow\infty$. This is the large-volume limit we are
interested in. For $L\m\rightarrow\infty$ one thus  obtains in
$D=4-2\e$ dimensions, 
\eq
V_{1-loop}= \sum_{1\leq i<j\leq n} \mu^D v_{ij}^2\left[\frac{(4\pi)^2\m^{2\e}
}{n\a\d^2 g^2} 
-2 v_{ij}^{-\e}
\cos(\half\pi\e)\G(\e-2)\right] 
\eqn{VD}

The $1/\e$-term in the expansion of\equ{VD} is canceled  by a
counter term $(1-Z)L^D \tr {\gb^2\over\a g^2}$ in the
classical action. To order $\hbar$, we require in the MS-scheme,
\eq
Z=1+ \frac{n \d^2\a g^2\m^{-2\e}}{16\pi^2\e}
\eqn{counter}

Introducing the scale $\k$ 
\eq
\ln\frac{\k^2}{ 4\pi\mu^2}=-\frac{(4\pi)^2}{ n\a \d^2
g^2}-\g_E+1+O(\ln g^2,g^2), 
\eqn{a}
the expression\equ{VD} for the 1-loop effective potential in 
$D=4$ dimensions can be written
\eq 
V_{1-loop}=\sum_{1\leq i<j\leq n}\frac{(\lm_i-\lm_j)^2\d^2}{
32\pi^2}\ln\left[\frac{(\lm_i-\lm_j)^2\d^2}{ e\k^4}\right]     
\eqn{EFF}
which is \equ{veff} in section~4.

The counter term  of  the  classical potential together with $Z_\a$
and $Z_g$ given in\equ{zfac}, also determines the renormalization
constant $Z_{\gb^2}$ of the global ghost $\gb$ to order
$g^2$. Using\equ{counter} we obtain in the MS-scheme  
\eq
Z_{\gb^2}=Z Z_\a Z_g^2=1-\frac{n g^2\m^{-2\e}}{32\pi^2\e}(3+\a(1-2\d))
+O(g^4)
\eqn{Zgb2}
Note that a derivation of the critical exponent of
$\gb$  via the effective potential only involves
UV-divergences and that IR-divergences are absent from this calculation.

\section{One-loop contribution to the vertex\equ{vertex}}
We are interested in the color antisymmetric part of the  mixing
between the composite field $\OO^a$ given in\equ{Odef} and $\pa\ipr A$
for vanishing momentum transfer.
The one-loop contribution to the vertex\equ{vertex} is obtained by
evaluating  the infinite sum of ghost loop diagrams of Fig.~5 
\vskip .5cm
\hskip 1cm\psfig{figure=vertexloop.ps,height=1.3in}
\nobreak\newline
{\small\baselineskip 5pt 
\noindent Fig.~5: Nonvanishing one-loop contribution to the
vertex\equ{vertex}.} 

Contributions antisymmetric in the
color indices  arise only from diagrams with an odd number of
insertions of the matrix $\hat\gb_{ab}=f^{abc}\gb^c$. For simplicity,
we evaluate this contribution to the vertex for an $SU(2)$ gauge
group. In a parametric form suitable
for performing the loop integral, the effective ghost propagator in
momentum space is
\eqa
G_{ab}(p^2)&\equiv&\frac{1}{p^2\one +\d\hat\gb}=\int_0^\infty d\lm
e^{-\lm (p^2\one + \d\hat\gb)}\cr
&=& \int_0^\infty d\lm
e^{-\lm p^2}\sum_{j=0}^\infty \frac{(-\d\lm\hat\gb)^j}{j!}\cr
&=& \int_0^\infty d\lm
e^{-\lm p^2}\left[\frac{\d^2\gb^a\gb^b}{\k^4} +\gb_\perp^{ab} \cos(\lm\k^2)
+\d\frac{\e^{abc}\gb^c}{\k^2}\sin(\lm \k^2)\right]
\eqan{ghprop}
where
\eq
\gb_\perp^{ab}
=\d^{ab}-\d^2\frac{\gb^a\gb^b}{\k^4}=-\d^2\frac{\hat\gb_{ac}\hat\gb_{cb}}{\k^4}
\eqn{trans}
projects on color degrees of freedom transverse to $\gb$ and 
$\k^4=\d^2\gb^a\gb^a$ is the scale of SSB of the effective
potential\equ{veff} for an $SU(2)$ theory. The fact that the summation
in \equ{ghprop} gives  a relatively simple expression for an $SU(2)$
group, greatly facilitates the  
evaluation of the one-loop contribution to the vertex. Note that the
poles of the Euclidean ghost propagator\equ{ghprop} are all at purely imaginary
$p^2$  and that the massless ghost is precisely that
of the unbroken $U(1)$. The sum of Feynman diagrams  with $p$
insertions of $\hat\gb$  in $D$
space-time dimensions corresponds to the loop-integral 
\eq
\G^{ab}_\m(q)=i q_\m \G^{ab}_{\OO\,\pa\ipr A}(q^2)= -ig^2\a\d\intd{k} \e^{cad} G_{de}(k^2)\e^{ebg}
G_{gc}((k-q)^2)) (\d k_\m +(1-\d) (k-q)_\m)
\eqn{v1}
Using the parameterization\equ{ghprop} for the ghost propagator,
performing the color summations and retaining only the part of $\G^{ab}_{\OO\,\pa\ipr A}$
that is antisymmetric in the color indices one obtains from\equ{v1}
\eq
\G^{[ab]}_{\OO\,\pa\ipr
A}(q^2)=g^2\a(2\d-1)\d^2\frac{\e^{abc}\gb^c}{\k^2}
\intf{k}\int_0^\infty d\lm \sin(\lm\k^2)\int_0^\infty d\lm' e^{-\lm
k^2-\lm'(k-q)^2}
\eqn{v2}
where we have already taken the limit $D\rightarrow 4$ since the
integral\equ{v2} is UV and IR finite. The vertex in the limit of
vanishing momentum transfer 
$q^2\rightarrow 0$ is particularly simple to evaluate
\eqa
\G^{[ab]}_{\OO\,\pa\ipr
A}(q^2\rightarrow 0)&=&\e^{abc}\gb^c \frac{g^2\a(2\d-1)\d^2}{16\pi^2\k^2}
\int_0^\infty d\lm \frac{\sin(\k^2\lm)}{\lm}\cr
&=&\e^{abc}\gb^c \frac{g^2\a(2\d-1)\d^2}{32\pi\k^2}
\eqan{v3}
Using that $\a(2\d-1)=3$ in CCG, the last expression in\equ{v3} for
the antisymmetric vertex at vanishing momentum transfer
becomes\equ{vertex} of the main text. 

\section{RG-invariance of the effective long-range interaction}
We here determine the renormalization of the 
effective interaction\equ{interac2} to order $g^2$ for planar contributions to
large Wilson loops of the kind shown in Fig.~2. Power counting
suggests that UV-divergences do not arise from this effective
interaction since the coupling $K$ has positive mass dimension. To
lowest order in $g(\m)$ the interaction\equ{interac2}
is just the product of the gluon propagator and the coupling
$g^2$. This interaction therefore depends on the
renormalization point $\m$. We will show that this $\m$-dependence of
the lowest order interaction is however canceled
to order $g^2$ by vertex corrections in the Wilson loop. 

The Wilson loop itself is a RG-invariant\cite{do80} observable,
and can therefore be expressed in terms of the bare coupling $g_B$ and
the bare gluonic field $A_B$ which do not depend on the
renormalization scale.
The bare field and coupling are related to the corresponding
renormalized quantities by\equ{rconst}, with the renormalization
constants of\equ{zfac}. The product of propagator and coupling
constant $g^2$ renormalizes with $Z_g^2 Z_3$
\eq
g_B^2\vev{A_B(x) A_B(0)}=Z_g^2 Z_3 g^2\vev{A(x) A(0)}
\eqn{renprop}
\equ{renprop} is proportional to the effective 2-point
interaction\equ{interac} in the Wilson loop however only to {\it lowest} order
in $g^2$. In order $g^4$ the divergent parts of the
diagrams in Fig.~6 also 
contribute to the effective 2-point interaction\equ{interac2}. As we
will see, these divergences compensate the factor 
$Z_g^2Z_3$ in\equ{renprop} to order $g^2$.  This implies that the (finite) 
corrections from the contributions of Fig.~6 precisely cancel
the $\m$ dependence of the renormalized effective 2-point interaction to
order $g^2$. The anomalous dimension of the effective 2-point
interaction\equ{interac2} is then
at least of order $g^4$, which is sufficient to prove that it is
renormalization group invariant in the limit $\m\rightarrow\infty$.

We show that the divergent corrections from planar diagrams of
order $g^2$ shown in Fig.~6 can be absorbed in the {\it longitudinal}
interaction by a multiplicative factor
\eq
(Z_g^2 Z_3)^{-1} =1 +\frac{ g^2n\m^{-2\e}}{32\pi^2\e}(3+\a) +O(g^4)
\eqn{Zinv}
if the lowest order interaction is longitudinal (i.e. we 
consider only $O(g^2)$ corrections to the long-range
interaction). 

Only the self-energy insertions (Fig.~6a) and vertex correction
 (Fig.~6b) are UV-divergent 
in dimensional regularization as $\e=(4-D)/2\rightarrow 0$. To
evaluate the divergences, only the behavior of propagators at high
 momenta are needed and perturbative propagators and vertices are therefore
 sufficient in an asymptotically free model.   
\vskip .5cm
\hskip 0.1cm\psfig{figure=wlcorr.ps,height=1.3in}
\nobreak\newline
{\small\baselineskip 5pt 
\noindent Fig.~6: Planar corrections of order $g^2$ to the effective
longitudinal interaction. Sections of the Wilson loop are shown in
bold and the path ordering is indicated by arrows; the effective
longitudinal interaction is depicted as a thin line and perturbative
gluon exchanges as wavy lines. {\it a)} A self-energy insertion on the
Wilson line. {\it b)} A perturbative vertex corrections to the long-range
longitudinal gluon interaction. {\it c)} Counter-term $Z_g^2 Z_3-1$ due
to renormalization of the effective longitudinal interaction. {\it d)}
A finite perturbative correction of order $g^2$.} 

Let us first analyze the vertex correction of Fig.~6b). In the Wilson
loop this correction gives rise to a loop integral of the form 
\eqa
\left.p_\o\Gamma_\o^a(p)\right|_{Fig.6b}&=&-ig^2\frac{n}{2}t^a\int\int
d\t\,d\t^\prime 
\frac{dx_\m}{d\t} \frac{dy_\n}{d\t^\prime}\Theta(\t-\t^\prime)\intd{k}
e^{i(k(x-y)-py)}\times\cr &&\kern-5em\times
  D_{\m\r}(k)\left[\d_{\r\s}(k^2-(p+k)^2)-k_\r k_\s+(p+k)_\r
(p+k)_\s\right] D_{\s\n}(p+k)
\eqan{vert1}  
where we have used that the lowest order effective interaction is
longitudinal. In\equ{vert1} the color traces have been
performed, $D_\mn(p)$   denotes the perturbative
gluon propagator and $x(\t), y(\t^\prime)$ are points on the Wilson
loop. The $\Theta$-function in\equ{vert1} arises due to path ordering.
Note that the term in square brackets of\equ{vert1} is a linear
superposition of transverse projectors. Inserting the explicit
expression for the perturbative gluon propagator, the integrand
of\equ{vert1} can be rewritten as
\eqa
D_{\m\r}(k)\left[\d_{\r\s}(k^2-(p+k)^2)-k_\r k_\s+(p+k)_\r
(p+k)_\s\right] D_{\s\n}(p+k) = D_\mn(p+k)-\cr 
- D_\mn(k)+\frac{k_\m p_\n + p_\m (p+k)_\n}{(p+k)^2 k^2}+(\a-1)(k_\m
(p+k)_\n  (k^2+pk) \frac{(k^2+pk)(p^2+2pk)}{k^4(p+k)^4}\cr   
\eqan{vert2}
The first two terms in the last expression for the integrand
of\equ{vert1} precisely compensate the divergent contributions from
the self-energy insertions\footnote{Note that the 
long-range interaction in Fig.~6a) is longitudinal -- $\m$-dependent
divergences arise only from so-called 
``pinch'' contributions\cite{co81} with at least two coinciding
vertices. Apart from a difference in the overall sign, the integrands of
these ``pinched'' self-energy contributions 
are just the first two terms in\equ{vert2}.} of Fig.~6a). The
other terms in\equ{vert2} are 
longitudinal in one or the other (or both) gluon vertices. The
$\m$-dependent divergent contributions of order $g^2$ from the
vertex- {\it and} self-energy corrections to the effective 
interaction therefore arise from the divergent part of  the integral
\eqa
\left.p_\o\Gamma_\o^a(p)\right|_{Fig.6a+b}&=& {\rm convergent} + \frac{g^2n}{2}
t^a\int\int d\t\,d\t^\prime\,\Theta(\t-\t^\prime) \frac{dx_\m}{d\t}
\frac{dy_\n}{d\t^\prime}\times\cr
&&\kern-10em\times\intd{k}\left[ \frac{1}{(k-p/2)^2(k+p/2)^2}
\left(p_\m\pad{}{y_\n} -p_\n\pad{}{x_\m}\right)+\right.\cr
&&\kern-10em\left.+(\a-1)\frac{pk(k^2-p^2/4)}{(k+p/2)^4(k-p/2)^4}
\left((k-p/2)_\m\pad{}{y_\n}
-(k+p/2)_\n\pad{}{x_\m}\right)\right]e^{i[(k-p/2)x-(k+p/2)y]}\cr  
\eqan{vert3} 
Observing that the remaining integral
is divergent only for $x=y$, one can now perform one of the line integrals
and obtains 
\eqa
\left.p_\o\Gamma_\o^a(p)\right|_{Fig.6a+b}&=& {\rm convergent}+ g^2n
t^a\int d\t\, \frac{dx_\m}{d\t} e^{-ipx} p_\m \times\cr
&&\times\intd{k}\left[ \frac{1}{(k-p/2)^2(k+p/2)^2}
+\frac{\a-1}{D}\frac{k^2 (k^2-p^2/4)}{(k+p/2)^4(k-p/2)^4}\right]\cr
&=& {\rm convergent}+\frac{g^2 \m^{-2\e} n }{64\pi^2\e}(3+\a)t^a\int d\t\,
\frac{dx_\m}{d\t} e^{-ipx} p_\m
\eqan{vert4}  
as the divergent part of the vertex and self-energy
corrections to the effective longitudinal interaction
\footnote{Note that for a {\it transverse} gluonic interaction the
divergent  perturbative
corrections could not be absorbed multiplicatively, since
the operators $A_{[\m,\n]}$ and $[A_\m,A_\n]$ mix in order
$g^2$.} in order $g^2$.  The divergent contributions we calculated
arise at both vertices of the effective interaction and thus
precisely cancel the counter-term $Z_g^2 Z_3-1$ from the renormalization of the
lowest order effective interaction to order $g^2$. In the Wilson loop, the
$\m$-dependence of the effective longitudinal interaction is therefore
compensated by the $\m$-dependence of  vertex and self-energy
corrections to order $g^2$.  

Only the perturbative diagrams considered here are planar and UV-divergent to
order $g^2$. It might seem that IR-divergent contributions arise from
the exchange of longitudinal perturbative gluons shown in Fig.~6d).
This is however not the case and the IR-divergences in fact cancel.    
To order $g^2$ the effective longitudinal 2-point interaction 
is independent of $\m$ and its  anomalous dimension is therefore of order
$g^4$. Perturbative 
corrections to the effective long range interaction are therefore analytic
in $g$ and $K$ and $s$ in\equ{interac2} approach
RG-invariant constants as $g\rightarrow 0$.

\section{Integral equation for a planar Wilson loop with longitudinal
  interaction} 
In the following we derive the integral equation\equ{eqWR} satisfied by
a Wilson loop with an effective longitudinal gluonic 2-point
interaction\equ{interac2}. As 
has been argued in section~6, we need to consider only contributions
which are of leading order in an $1/n$ expansion -- for color-vector
interactions, these are planar diagrams with all exchanges inside the
loop. 

Consider a typical  contribution to the Wilson loop with only
2-point interactions of this type. We open the loop at some point $A$
as indicated  in Fig.~7. Since the vertices on  the loop 
are {\it ordered} and the exchanges are planar in the above sense,
there is a one-to-one correspondence between one-sided planar
contractions of vertices on a ``line''  and corresponding 
contributions to the Wilson loop if each topologically different
one-sided planar diagram of the ``line'' is counted exactly
once. Although several topologically different diagrams on the
``line'' may correspond to the same diagram for the
Wilson loop, this multiplicity precisely accounts for the statistical
factor of that diagram in the loop. We are thus left with the
combinatorial  problem
of generating all topologically different one-sided planar
contributions to the ``line'' exactly once. 
\vskip .5cm
\hskip 1cm \psfig{figure=wl2.ps,height=1in}
\nobreak\newline
{\small\baselineskip 5pt 
\noindent Fig.~7: A planar contribution to the Wilson loop from a
2-point interaction and the corresponding one-sided planar
``line''-diagram when the loop is opened at $A$}

The closed curve of the Wilson loop 
$\CC:=\{x_\m(\t):\t\in[0,2\pi]{\rm\ with\ } x_\m(0)=x_\m(2\pi)\}$, is
parameterized by $\t\in[0,2\pi]$. In the following we reserve  Greek
letters for the one-dimensional coordinates on the
loop respectively ``line''. An effective longitudinal interaction of the form
$v_\mn^{ab}(x-y)=\d^{ab}\pa_\m\pa_\n \s((x-y)^2)/n$ in leading order
of $1/n$ results in a scalar factor 
\eq
v(\a,\b)=\pad{}{\a}\pad{}{\b} \s((x(\a)-y(\b))^2)
\eqn{vs}
for each interaction on the ``line'' between points $\a$ and $\b$ of
such a one-sided planar diagram, since the color matrices at each
vertex always contract to give the quadratic Casimir of the
fundamental representation. In leading order of $1/n$, the ``line'' is
therefore proportional to the unit matrix. Note that for a
{\em transverse} exchange the interaction depends on the 
{\em direction} relative to the line element and is not of the  
form\equ{vs}.  For an abelian group there is
no distinction between the ordering of the vertices and {\it any}
longitudinal interaction gives a vanishing
contribution to  the Wilson loop, since\equ{vs} is a total derivative in
the loop parameter. In non-abelian gauge theories, the ordering of the
vertices does however matter and longitudinal interactions do
contribute to the Wilson loop. Without
the spontaneous symmetry breaking, the contribution from 
longitudinal 2-point interactions would be compensated by
contributions from Green's functions of higher order. As explained in
the main text, we however do not expect such a compensation of the long range
longitudinal interaction  when the global $SU(n)$-symmetry is
spontaneously broken.  

Let us denote by $F_k(\a,\b)$ the sum of all one-sided planar diagrams
containing $k$ longitudinal interactions  between the points $\a$ and
$\b$ on the ``line'',
with each topologically distinct diagram counted exactly 
once. Diagrammatically, $F_{k+1}(\a,\b)$ is related to 
$F_l(\a,\b)$ with $l\leq k$ by:   
\vskip .5cm
`\hskip 1cm \psfig{figure=wl1.ps,height=1in}
\nobreak\newline
{\small\baselineskip 5pt 
\begin{center}Fig.~8: Iterative equation for $F_{k+1}(\a,\b)$.\end{center}}

Fig.~9 expresses the fact that the first interaction in a one-sided
planar diagram of order $k+1$ ``encloses'' a one-sided planar diagram of order
less than $k$ and that the remaining diagram after this
first interaction  is one-sided planar and of the complimentary order. 
Naturally, one has to sum over all intermediate orders and integrate
over all possible (ordered)
positions of the vertices for the first interaction between the points $\a$ and
$\b$. It is not hard to show inductively that each topologically distinct
one-sided planar diagram of order $k+1$ appears exactly once in
$F_{k+1}$. Written explicitly, the
one-sided planar contribution of order $k+1$ therefore is,
\eq
F_{k+1}(\a,\b)=\sum_{m=0}^k\int_\a^\b\,d\g\,
\int_\a^\g\,d\tau\,v(\tau,\g)F_m(\tau,\g)F_{k-m}(\g,\b)
\eqn{b1}
where $F_0(\a,\b)=1$ is the contribution with no
interactions.  The total one-sided planar contributions $W(\a,\b)$ to
the ``line'' between $\a$ and $\b\geq\a$  is just the sum of all $F_k$,
since each has the statistical factor~1:
\eq
W(\a,\b)=\sum_{k=0}^\infty\,F_k(\a,\b)=F_0(\a,\b) +
\sum_{k=0}^\infty\,F_{k+1}(\a,\b) 
\eqn{b2}
Using\equ{b1} and rearranging the double sum, one obtains
\eqa
W(\a,\b)&=&F_0(\a,\b)+\sum_{k=0}^\infty \sum_{m=0}^k \int_\a^\b\,d\g\,\int_\g^\b\,d\tau\,v(\tau,\g)F_m(\tau,\g)F_{k-m}(\g,\b)\nonumber\\
&=&1+\int_\a^\b\,d\g\,\int_\a^\g\,d\tau\,v(\tau,\g)W(\tau,\g)W(\g,\b)\nonumber\\&=&1+\int_\a^\b\,d\tau\,\int_\tau^\b\,d\g\,v(\tau,\g)W(\tau,\g)W(\g,\b).
\eqan{b3}

Relatively simple is the case of a circular Wilson loop
parameterized by an angle with a longitudinal interaction of the
form\equ{vs}.  The distance between points on the circular loop in
this case depends only on the difference in their angular positions
and the radius $R$:
$(x-y)^2=4R^2\sin^2\left(\frac{\t_x-\t_y}{2}\right)$. For a longitudinal
and translationally and rotationally invariant interaction,
$v(\a,\b)=v_R(\a-\b)$ of\equ{vs} in this case 
is a function of the angular difference $\a-\b$ and the radius
only. Thus, for a 
circle of radius $R$, $W(\a,\b)=W_R(\a-\b)$ also depends only on the angular
difference and one can rewrite \equ{b3} as:
\eqa
W_R(\t)&=&1+\int_0^\t\,d\t'\,\int_{\t'}^\t\,d\t''\,v_R(\t''-\t')W_R(\t''-\t')W_R(\t-\t'')\nonumber\\
&=&1+\int_0^\t\,d\t'\,\int_0^{\t'}\,d\t''\,v_R(\t'')W_R(\t'')W_R(\t'-\t'').
\eqan{b4}
To solve\equ{b4} numerically, it is advantageous to consider instead
the equivalent non-linear integro-differential equation: 
\eq
\frac{d~}{d\t}W_R(\t)=\int_0^\t\,d\t'\,v_R(\t')W_R(\t')W_R(\t-\t'),
\eqn{b5}
with the initial value $W_R(0)=1$. 
The expectation value of the Wilson loop in this approximation is
then $nW_R(2\pi)$.
\end{appendix}

}\end{document}